\documentclass[twocolumn,superscriptaddress,floatfix,showpacs]{revtex4-1}
\pdfoutput=1
\usepackage{graphics,amssymb,amsmath,epsfig,color}
\usepackage{graphicx}

\newcommand{\be}{\begin{equation}} \newcommand{\ee}{\end{equation}}
\newcommand{\bea}{\begin{eqnarray}} \newcommand{\eea}{\end{eqnarray}}
\begin{document}

\title{SIR epidemics with long range infection in one dimension}
\author{Peter Grassberger} \affiliation{JSC, FZ J\"ulich, D-52425 J\"ulich, Germany}
\date{\today}

\begin{abstract}

We study epidemic processes with immunization on very large 1-dimensional lattices,
where at least some of the infections are non-local, with rates decaying as power laws
$p(x)\sim x^{-\sigma-1}$ for large distances $x$. When starting with a single infected 
site, the cluster of infected sites stays always bounded if $\sigma >1$ (and dies with
probability 1, if its size is allowed to fluctuate down to zero), but the process can 
lead to an infinite epidemic for $\sigma <1$. For $\sigma <0$ the behavior is essentially 
of mean field type, but for $0<\sigma\leq 1$ the behavior is non-trivial, both for the 
critical and for supercritical cases. For critical epidemics we confirm a previous prediction 
that the critical exponents controlling the correlation time and the correlation length
are simply related to each other,
and we verify detailed field theoretic predictions for $\sigma \searrow 1/3$.
For $\sigma =1$ we find generic power 
laws with continuously varying exponents even in the supercritical case, and confirm in detail 
the predicted Kosterlitz-Thouless nature of the transition. Finally, the mass $N(t)$ 
of supercritical clusters grows for $0<\sigma < 1$ like a stretched exponential. 
This implies that networks embedded in $1$-d space with power-behaved link distributions 
have infinite intrinsic dimension (based on the graph distance), but are not small world. 

\end{abstract}

\maketitle

\section{Introduction}

Epidemic spreading, both on regular lattices and on random graphs, have attracted 
increasing attention in the statistical physics community \cite{Grass83,Newman02}. 
Basically one distinguishes between two 
types of epidemics. In both all agents except for the {\it seeds} are susceptible,
while infectious agents (such as the seeds in the initial configuration) stay infectious 
for a finite time (during one time step in the following), which means that they can 
infect agents that share with them a link. After this 
infective period, they either become susceptible again (SIS model), or they become
``removed", i.e. either immune or dead (SIR model). In both cases, agents stay at fixed
places (whence they can be identified with lattice sites or network vertices), so that 
the entire dynamics is contained in the changes of their internal (S, I, R) states.

There exist of course several generalizations of these two basic schemes (e.g. 
cooperative infection \cite{Janssen04,Bizhani}, infection using the ``power of 
choice" \cite{Achlioptas}, moving agents \cite{Wijland}, memory effects \cite{Chate}, 
etc.), but even the basic schemes
show rich behavior, if we allow for different network topologies. 

In the following we shall only deal with basic SIR epidemics, with discrete time and 
infective period equal to one time step. In this case, the process must
die out on any finite system, since susceptibles are used up but not replenished, and 
the set of ``removed" sites becomes, for large times, just a percolation cluster.
More precisely, we shall discuss two models, both of which involve infection over 
large distance, with the infection probability decaying as some inverse power of the 
distance \cite{Grassberger86,Janssen99,Linder}. In both, the sites can be viewed as 
sites in an infinitely large 1-d lattice (in the simulations we use $L=2^{64}$ sites, 
which is big enough so that we never have to worry about finite size effects, except 
in a few cases pointed out later \footnote{Such a large lattice cannot of course be 
stored explicitly in the computer memory. Instead we used hashing, implemented using 
linked lists, with a hash function $h(i) \equiv i \mod 2^m$ with $m \approx 20 - 30$. 
For a previous implementation and details, see e.g. \cite{percol-high_d}.}).
In both models, each infected site first attempts to infect $k_0$ other sites 
(and succeeds so with probability 1, if these sites are still susceptible), and then 
attempts with probability $p$ to infect one more site. Thus, the average number 
of newly attempted infections per node, which is also equal to the average out-degree of 
the generated graph, is 
\be
   k_{\rm out} = k_0 + p\;.
\ee
But both models differ slightly in how these sites
are connected, i.e. how the attempted infections are chosen:

\begin{itemize}
\item In model (A) we assume that in each attempt the target site is chosen randomly, 
with a distance $\pm x$ from the infectious site that is distributed according to 
a power law, 
\be
   P(x) \sim x^{-\sigma-1} 
\ee
for large $x$. More precisely, this distribution is obtained by first drawing a 
real-valued random number $y$ uniformly from the interval $(1/L^\sigma,1]$ (for $\sigma > 0$), 
 $[1,1/L^\sigma)$ (for $\sigma<0$) or $[0,\ln L]$ (for $\sigma=0$)  and setting
\be
   x =  \left\{
            \begin{array}{rl}
                  \lfloor y^{-1/\sigma}\rfloor   & \text{ for } \sigma \neq 0,\\
                  \lfloor e^y\rfloor        & \text{ for } \sigma =0.\\
            \end{array} \right.
\label{a-scaling}
\ee
Notice that we do {\it not} check that all attempts 
try to infect different targets. If several attempts are aimed at the same target,
all except the first one are simply lost.

\item In model (B) we first infect the left and right neighbors, and only in the 
subsequent infections distant sites are chosen, again with the same probability 
$P(x)$ given above.
\end{itemize}

In model (A), it can happen that {\it all} attempts during one time step try to 
infect target sites that are no longer susceptible, in which case the epidemic dies.
This cannot happen in model (B). There, the right neighbor of the rightmost infected
site is always susceptible, as is also the left neighbor of the leftmost site.
Model (B) is indeed, as far as the geometric structures obtained for $t\to\infty$
are concerned, a modification of the Watts-Strogatz \cite{WS} small world model that 
was studied previously in \cite{Kleinberg,Benjamini-2001,Coppersmith,Moukarzel,Sen,Juhasz-2012}.

In the next section, we will discuss the most important features common to both 
models. In Sec. III we shall treat in more detail the case $\sigma = 1$, where 
we find a number of non-trivial exact results. The critical case of model (A) 
with $0<\sigma <1$ is studied in Sec. IV [model (B) is always supercritical, as 
it never can die, thus no critical phase exists]. The supercritical case for 
$0<\sigma <1$ is finally discussed in Sec. V. The paper concludes with a discussion
in Sec. VI.

\section{General features}

We always start with a single seed located at the origin, $x=0$. Boundary conditions
are periodic. The jump probability $P(x)$ is cut off at $x=L$, i.e. the length of 
the widest jumps allowed is precisely the lattice size. The distribution 
of infected (or ``active", as we shall call them in the following) sites at time 
$t\geq 0$ is denoted as $\rho(x,t)$. The distribution of removed sites is then
\be
   R(x,t) = \sum_{t'=0}^{t-1} \rho(x,t').
\ee
The average number of active sites at time $t$ is then
\be
   n(t) = \sum_x \rho(x,t),
\ee
while the number of immune sites is 
\be
   N(t) = \sum_x R(x,t) = \sum_{t'=0}^{t-1} n(t').
\ee
Notice that $N(t)+n(t)$ can also be interpreted as the average number of sites
reached by chains of at most $t$ links from a randomly chosen pivot. Therefore,
if $N(t) \sim t^D$, the exponent $D$ would be the intrinsic (or ``topological"
\cite{Emmerich}\footnote{We avoid this name, in order to avoid confusion with 
the well established topological dimension discussed, e.g. in \cite{Menger}.})
dimension of the final network.

For model (B), $R(x,t)=1$ and $\rho(x,t)=0$ for all $|x|<t$. Thus the active sites
form two outgoing waves, one moving to the right and the other to the left. For 
$\sigma >0$ the infection is sufficiently short ranged that these two waves 
don't interfere with each other, in the limit $t\to\infty$. Basically the same 
is also true for model (A), although in that case $R(x,t)$ and $\rho(x,t)$ are 
no longer strictly 1 and 0, respectively, for $|x|<t$. Nevertheless, activity 
dies out also there for any finite $x$, i.e. $\rho(x,t)\to 0$ for $t\to\infty$.

\begin{figure}
\includegraphics[width=0.55\textwidth]{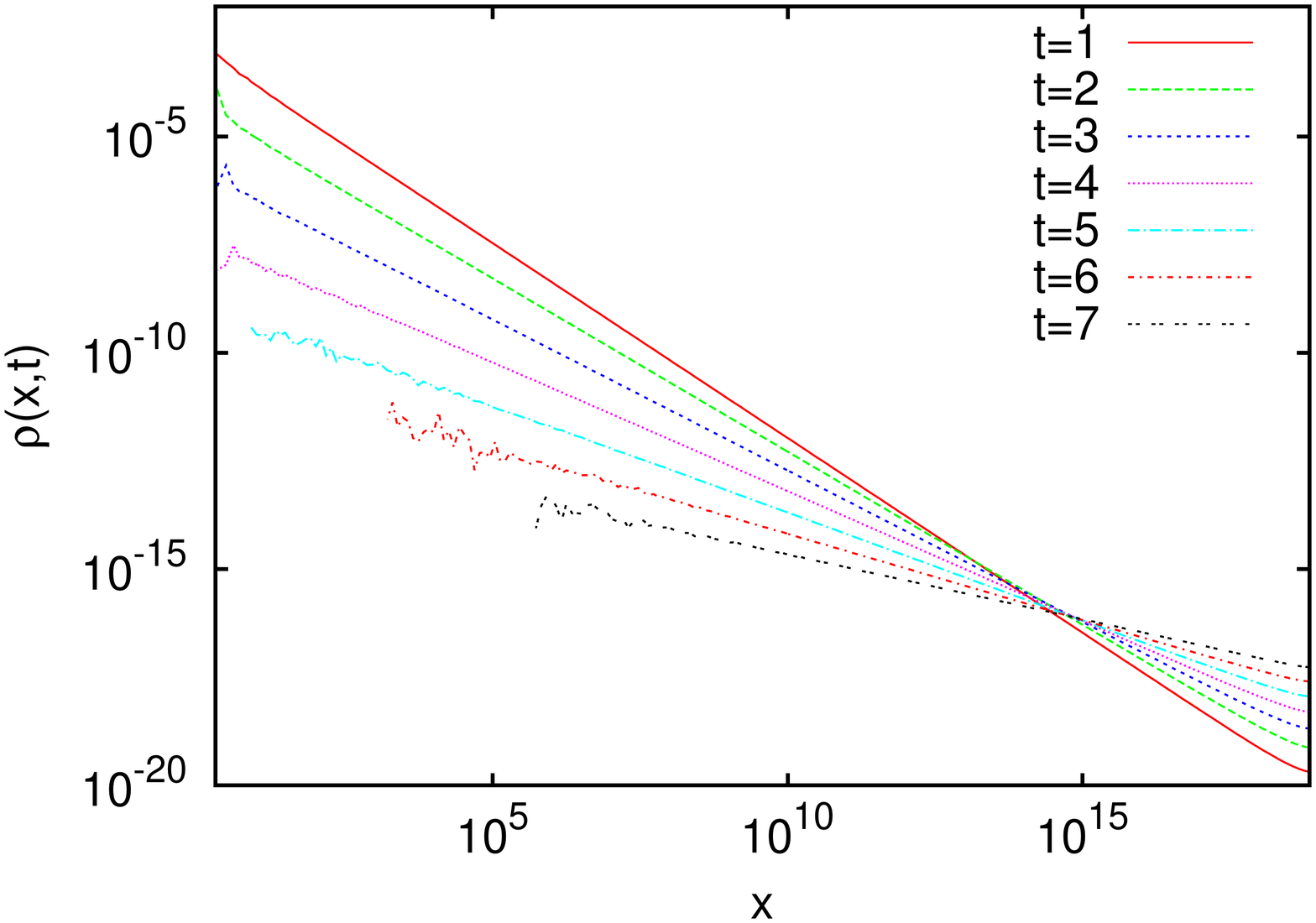}
\caption{(Color online) Densities of active sites at times $t=1,2,\ldots 7$ for 
model (A) with $\sigma=-0.1,\; k_0=2$, and $p=1$, i.e. with exactly 3 offsprings per active 
site. Because of left-right symmetry, we show here and in the following plots only 
the distribution for $x>0$. The densities become more and more uniform for increasing 
$t$, showing the mean field nature when $\sigma <0$.}
   \label{mf-active}
\end{figure}

This scenario no longer holds for $\sigma <0$ (for the case $\sigma=0$ see \cite{Coppersmith}). 
In that case, $P(x)$ would not be
normalizable for infinitely large lattices, hence finite lattice effects become 
important. Indeed, on large but finite lattices, both $R(x,t)$ and $\rho(x,t)$ 
become uniform, see Fig.~\ref{mf-active}. Also, the formation of finite loops is suppressed
by powers of $L$ (the number of loops with $m$ links, each limited in length to 
less than some constant, scales as $L^{1+m\sigma}$), so that the cluster 
of immunes becomes locally tree like in the limit $L\to\infty$ and the model 
becomes of mean field type.

\begin{figure}
\includegraphics[width=0.55\textwidth]{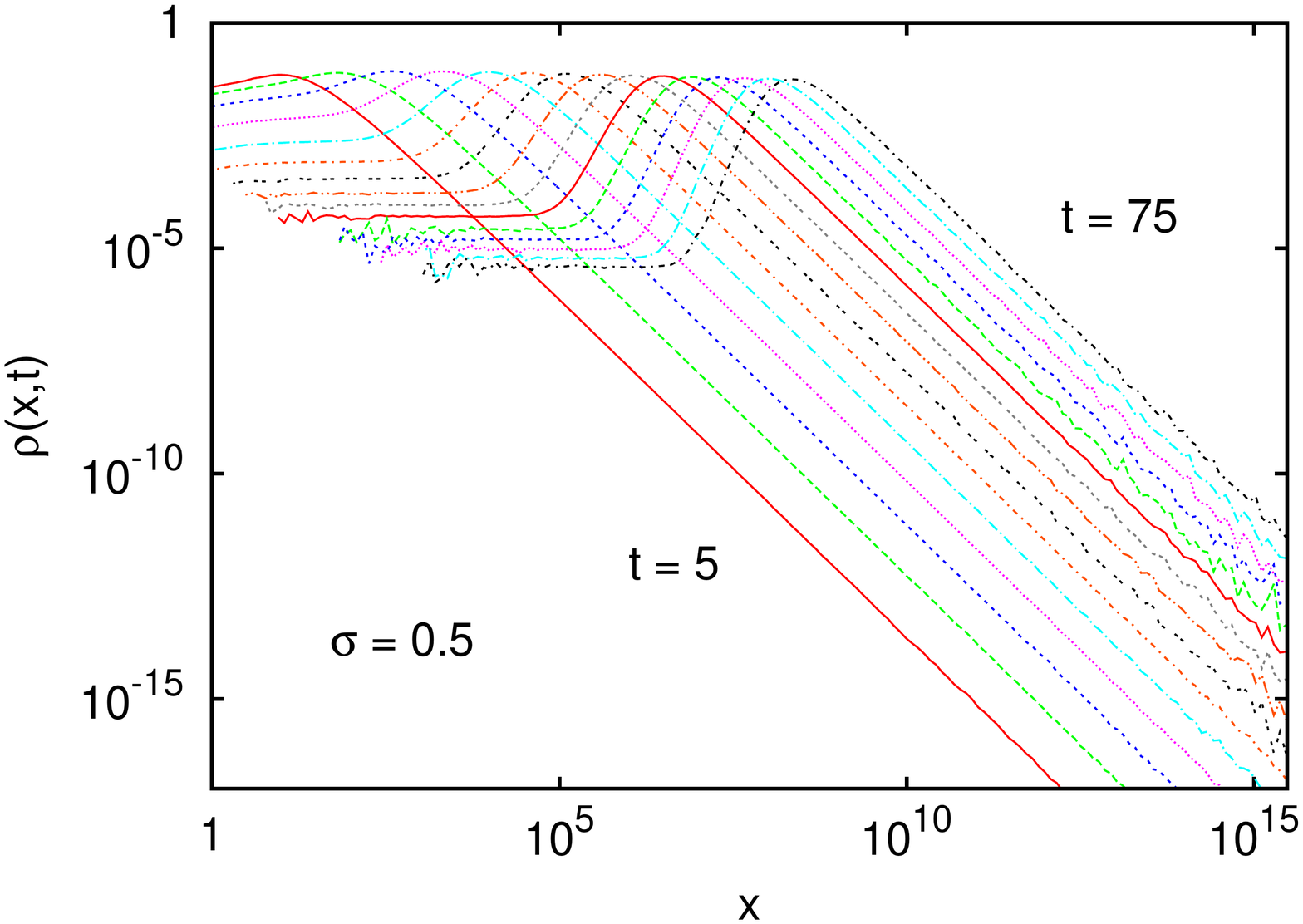}
\caption{(Color online) Densities of active sites at times $t=5,10,\ldots 75$ for
model (A) with $\sigma=0.5,\; k_0=1$, and $p=1$, i.e. with exactly 2 offsprings per active
site. Notice the tails decaying $~x^{-\sigma-1}$ for all times. For small $x$ the activity
dies out because (i) the density of immune sites gets higher and higher, and (ii) the 
parent activity shifts more and more outside.}
   \label{active-s15}
\end{figure}

\begin{figure}
\includegraphics[width=0.55\textwidth]{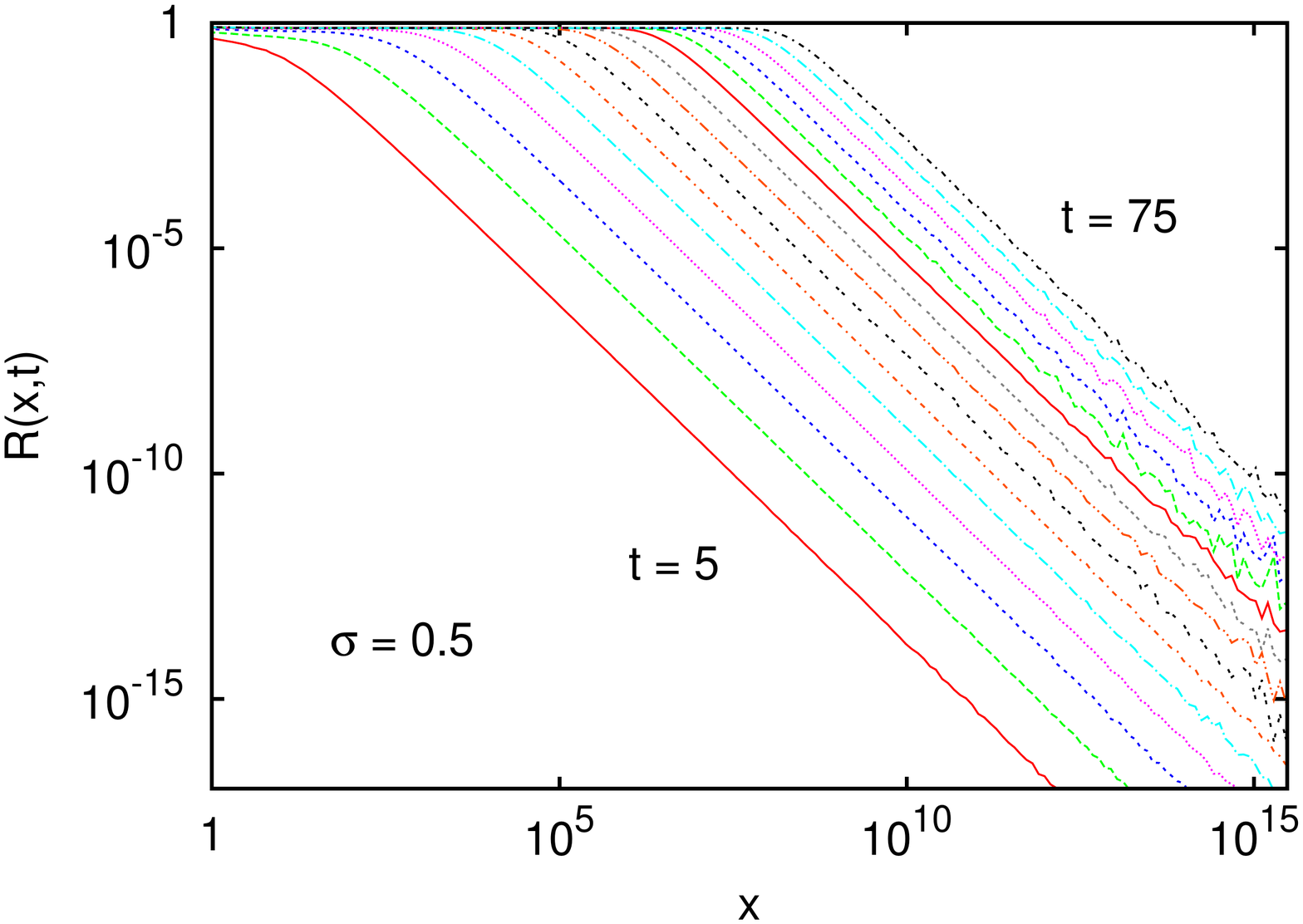}
\caption{(Color online) Densities of immune sites at at the same times as in Fig.~\ref{active-s15}.}
   \label{immune-s15}
\end{figure}

In contrast, for $\sigma>0$ it is found that both $R(x,t)$ and $\rho(x,t)$ decay
asymptotically as $|x|^{-\sigma-1}$ for all times $t\geq 0$, see e.g. 
Figs.~\ref{active-s15}, \ref{immune-s15} (see also \cite{Mancinelli}).
This is easily understood. First of all, they cannot decay faster, since the 
offspring distribution of a population concentrated at the origin would decay 
like that and any smearing due to a finite extend of the parent population 
can only make the offspring distribution wider. On the other hand, if $\rho(x,t)$
decays at some given time $t$ not slower than $|x|^{-\sigma-1}$, then the distribution 
of its offsprings is given for sufficiently large $|x|$ by 
\be
   \rho(x,t+1) \approx \sum_y \rho(y,t) P(x-y) \sim |x|^{-\sigma-1}.
\ee
In this expression we have neglected saturation effects (not all infections are 
successful, because not all sites are susceptible), but this approximation should 
be correct for large $|x|$ where most of the sites {\it are} susceptible. Since 
$R(x,t)$ is just a sum over $\rho(x,t')$ with $t'<t$, it satisfies the same asymptotic 
behavior. 

\begin{figure}
\includegraphics[width=0.55\textwidth]{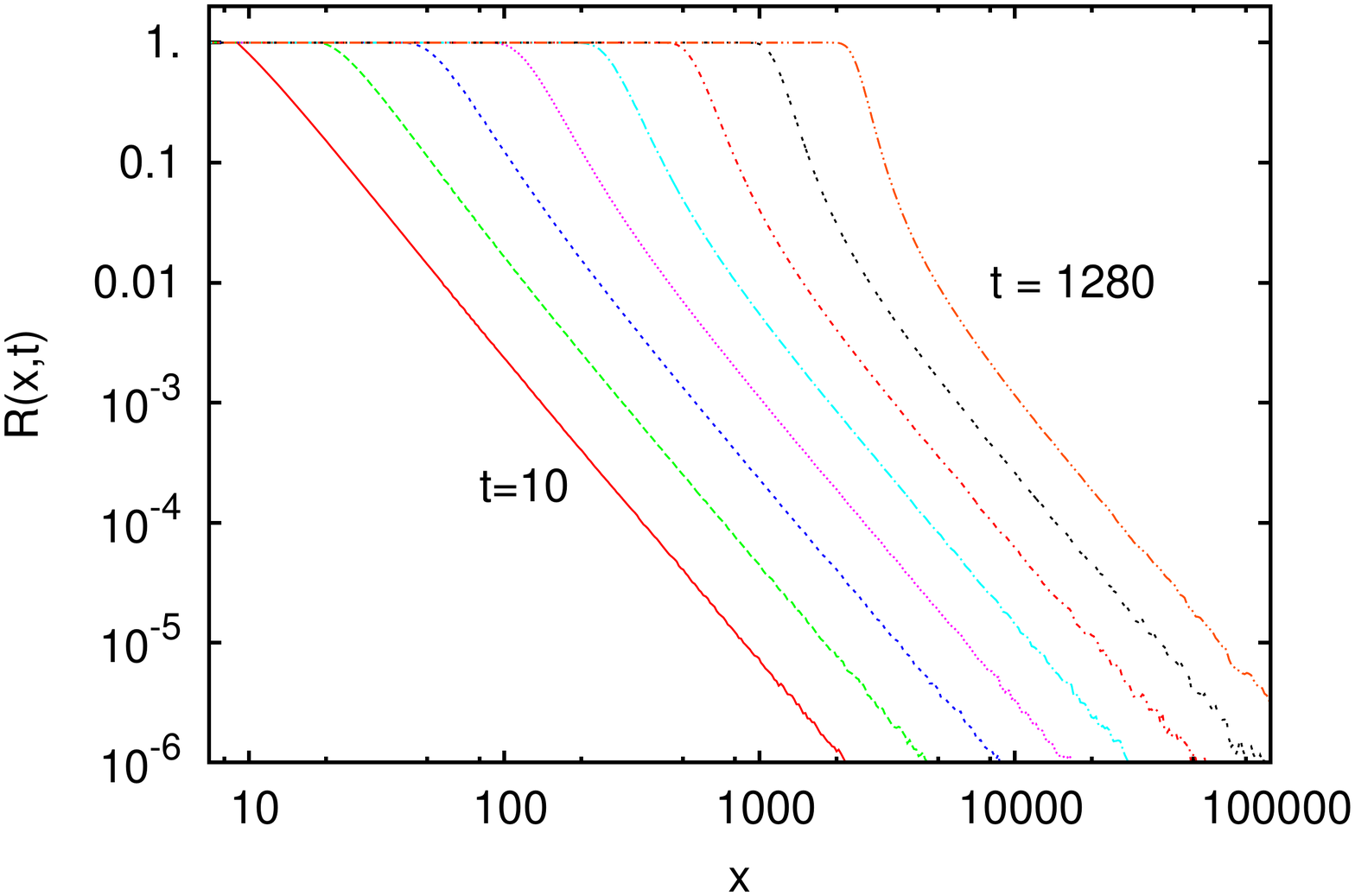}
\caption{(Color online) Densities of the immune cluster at times $t=10, 20, 40,\ldots 1280$ for 
model (B) with $\sigma=1.5, k_0=3$, and $p=1$, i.e. with two local and one long-range offsprings.
For large $x$, all densities decay $\sim 1/x^{\sigma+1}$, but for large times this tail sets in 
later and later.}
   \label{fig_B_subcrit}
\end{figure}

Notice that this argument only tells about the limit where we first let $x\to\infty$,
and then let $t$ become large. It does not prove that $R_\infty(x) = \lim_{t\to\infty}R(x,t)$
and $\rho_\infty(x) = \lim_{t\to\infty}\rho(x,t)$ decay asymptotically as $|x|^{-\sigma-1}$. For 
$0<\sigma \leq 1$ this seems to be correct nevertheless (see Figs.~\ref{active-s15},\ref{immune-s15}), 
but it it does not hold for $\sigma >1$. In that case the process dies with probability one for 
model (A) (for an elegant and simple proof, see \cite{Schulman}), but it survives forever for 
model (B). In the latter case the average number of active sites, $n(t) = \sum_x \rho(x,t)$,
tends to a constant and the wave of active sites has a stationary profile in a co-moving
frame (in a frame moving with constant velocity the profile widens due to fluctuations 
of the velocity). Profiles $R(x,t)$ are shown in Fig.~\ref{fig_B_subcrit} for model (B) with 
$\sigma=1.5$ and one long-range contact per site. We see that the tails decay $\sim |x|^{-\sigma-1}$ 
for all finite times, but that this behavior sets in later and later for increasing $t$. The 
bulk of the outgoing wave has finite width (i.e., becomes increasingly sharper in a log-log 
plot such as Fig.~\ref{fig_B_subcrit} ). In the terminology of \cite{Ebert-2000}, the fronts 
for model (B) are {\it pushed} when $\sigma >1$. 

The same is true for model (A), although one has to condition on 
(exponentially rare) surviving events in order to see this. This could be done by 
using e.g. PERM \cite{PERM}, but we have not done it.

\begin{figure}
\includegraphics[width=0.55\textwidth]{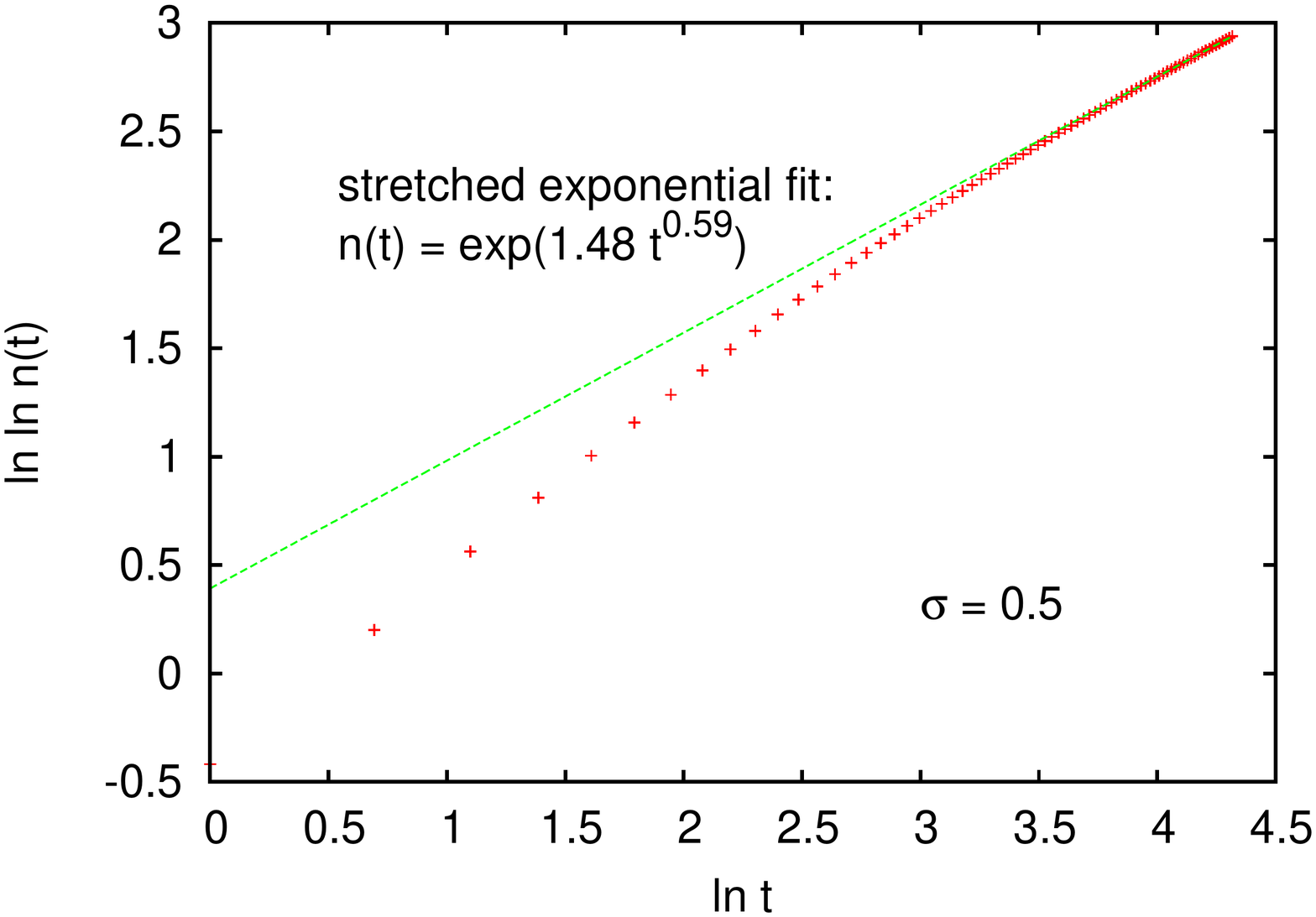}
\caption{(Color online) Plot of $\log \log n(t)$ versus $\log t$, where $n(t)$ is the average 
number of active sites, for the same run shown also in Figs.~\ref{active-s15} and \ref{immune-s15}.
The straight line represents a stretched exponential fit to the data with $50 \leq t \leq 75$, 
corresponding to $2.6\times 10^6 \leq n(t) \leq 1.6\times 10^8$.
Statistical errors are much smaller than the symbols.}
   \label{n_t-s15}
\end{figure}

Notice that the times in Fig.~\ref{fig_B_subcrit} are exponentially increasing. The linear progression 
of the front in the log-log plot then means that the immune cluster travels at constant speed. This 
is in contrast to the case $\sigma <1$ shown in Figs.~\ref{active-s15} and \ref{immune-s15} (where 
time increases linearly between successive curves). There the wave of infection travels with 
a speed that increases faster than a power with time, in agreement with \cite{Mancinelli,Castillo} 
but in contrast to \cite{Brockmann}). On the other hand, the decreasing distances between successive 
curves in Figs.~\ref{active-s15} and \ref{immune-s15} show that this increase of speed is less 
than exponential, in contrast to \cite{Mancinelli,Castillo}. In order to see this more clearly,
we plot $n(t)$ in Fig.~\ref{n_t-s15} for the runs shown in Figs.~\ref{active-s15}, \ref{immune-s15}. 
More precisely, Fig.~\ref{n_t-s15} shows $\log \log n(t)$ plotted against $\log t$. An exponential 
$n(t)\sim \exp(at)$ would correspond in this plot to a straight line with slope 1. This is obviously 
not observed (statistical errors are smaller than the symbol sizes). Rather, the data for large $t$
suggest a stretched exponential $n(t)\sim \exp(at^\gamma)$ with $\gamma = 0.59(1)$. But we should 
be careful in accepting this fit as the true asymptotic scaling. First of all, the data in 
Fig.~\ref{active-s15} are slightly curved, even for the largest $t$, suggesting that this estimate 
of $\gamma$ is too high. Also, fitting stretched 
exponentials is notoriously fraught with uncertainties. The same behavior is also seen for 
model (B) (data will be shown later). It shows also that the mass of clusters with diameter $\ell$
increases slower than exponentially with $\ell$, in contrast to claims made in \cite{Moukarzel}
(exponential increase is of course seen in the mean field regime, $\sigma < 0$).
More details will be given in Sec.~V.

\section{The case $\sigma = 1$}   \label{sigma_one}

It is well known that the case of interacting Levy flights with $\sigma = 1$ is very special,
in particular in one dimension of space. This was first found by Dobrushin, Ruelle \cite{Ruelle},
and Dyson \cite{Dyson}
who showed that 1-d Ising models with long range interactions can only have a finite temperature
phase transition, if $\sigma \leq 1$. Very soon after this, Anderson et al.~\cite{Anderson}
and Thouless~\cite{Thouless} showed that for $\sigma = 1$ one not only does have a phase 
transition, but that this transition is similar to the Berezinskii-Kosterlitz-Thouless (BKT)
transition in displaying a region with generic power laws in the supercritical phase with 
$\sigma$-dependent exponents. As in the XY model, the reason is that configurations
can be described as organized by defects which interact with each other by an
attractive logarithmic potential. This argument was later extended by Cardy~\cite{Cardy} 
to Potts and other models. Due to the Fortuin-Kasteleyn \cite{FK} relationship between the Potts 
model and percolation~\cite{Cardy}, this applies also to percolation and thus also to SIR 
epidemics. But it seems that the consequences for the latter have never been worked out in detail, 
with one notable exception: It was shown in~\cite{Aizenman} that the percolation transition 
for $\sigma = 1$ is discontinuous in the sense that the order parameter (the density of 
infected sites for $t\to\infty$) jumps discontinuously, when $p$ is increased through the 
percolation threshold. This might seem contradictory to the claim of universal power laws
(which usually hold only at continuous phase transitions), but several similar ``hybrid" cases, 
where aspects typical of a first order transition coexist with 
aspects of a second order transition, have been found recently also in other 
contexts~\cite{Dorogovtsev,Bizhani}. 

\begin{figure}
\includegraphics[width=0.55\textwidth]{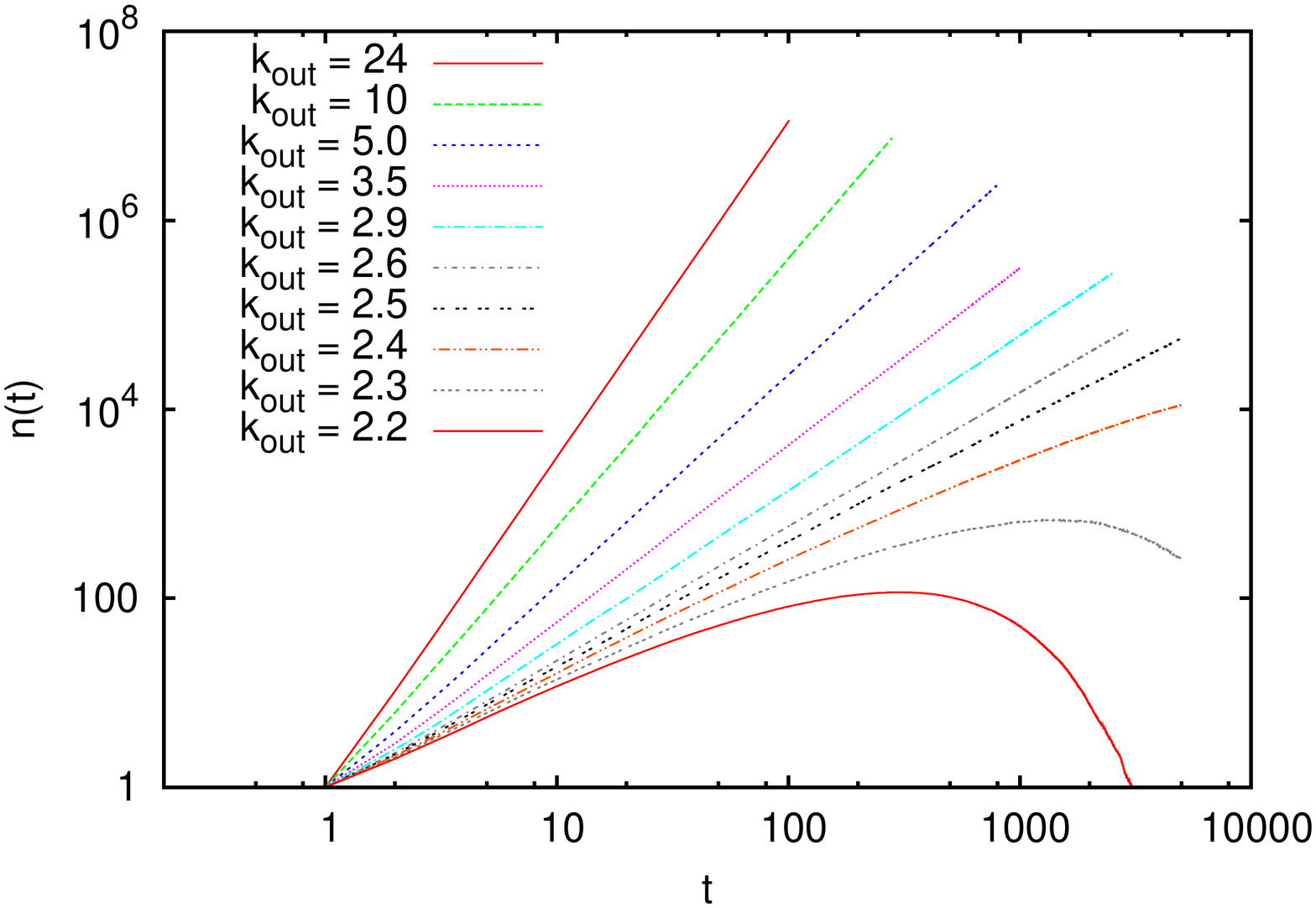}
\caption{(Color online) Log-log plot of $n(t)$ versus $t$ for model (A) with $\sigma=1$, 
and for different values of $k_{\rm out}$. All curves for $k_{\rm out}\geq 2.6$ seem to become 
asymptotically straight, giving a rough estimate of $k_{\rm out}= 2.6(1)$ for the critical point.}
   \label{n_t-s20}
\end{figure}

\begin{figure}
\includegraphics[width=0.55\textwidth]{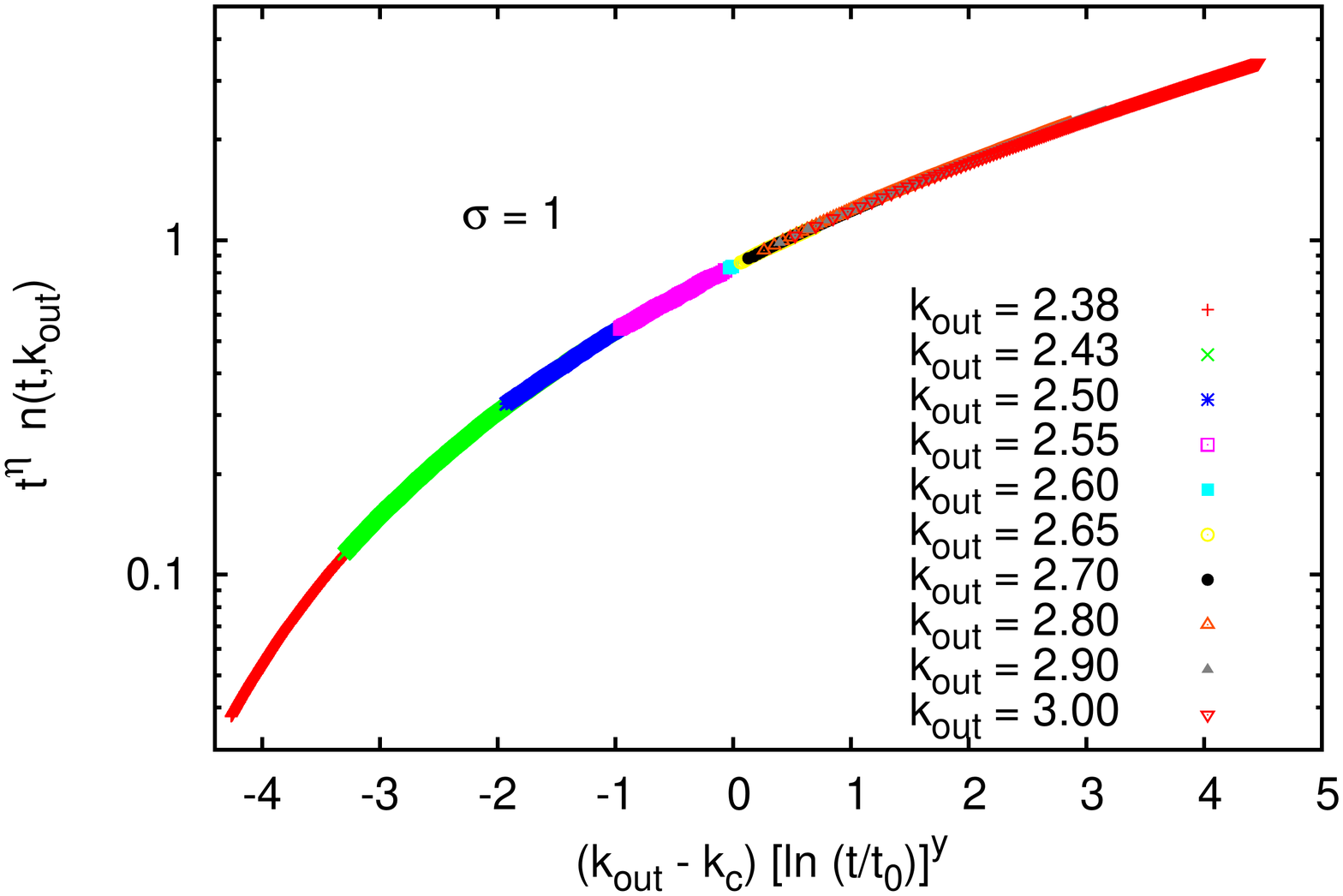}
\caption{(Color online) Data collapse for model (A), using the finite-$t$ scaling ansatz 
Eq.~(\ref{nut-log-scaling}). Parameters are $k_c = 2.602,\; \eta = 1.42,\; t_0 = 0.88,\;$ and 
$y = 1.37$. Only points with $t\geq 3$ are plotted.}
   \label{fig-BKT-fit}
\end{figure}

\begin{figure}
\includegraphics[width=0.55\textwidth]{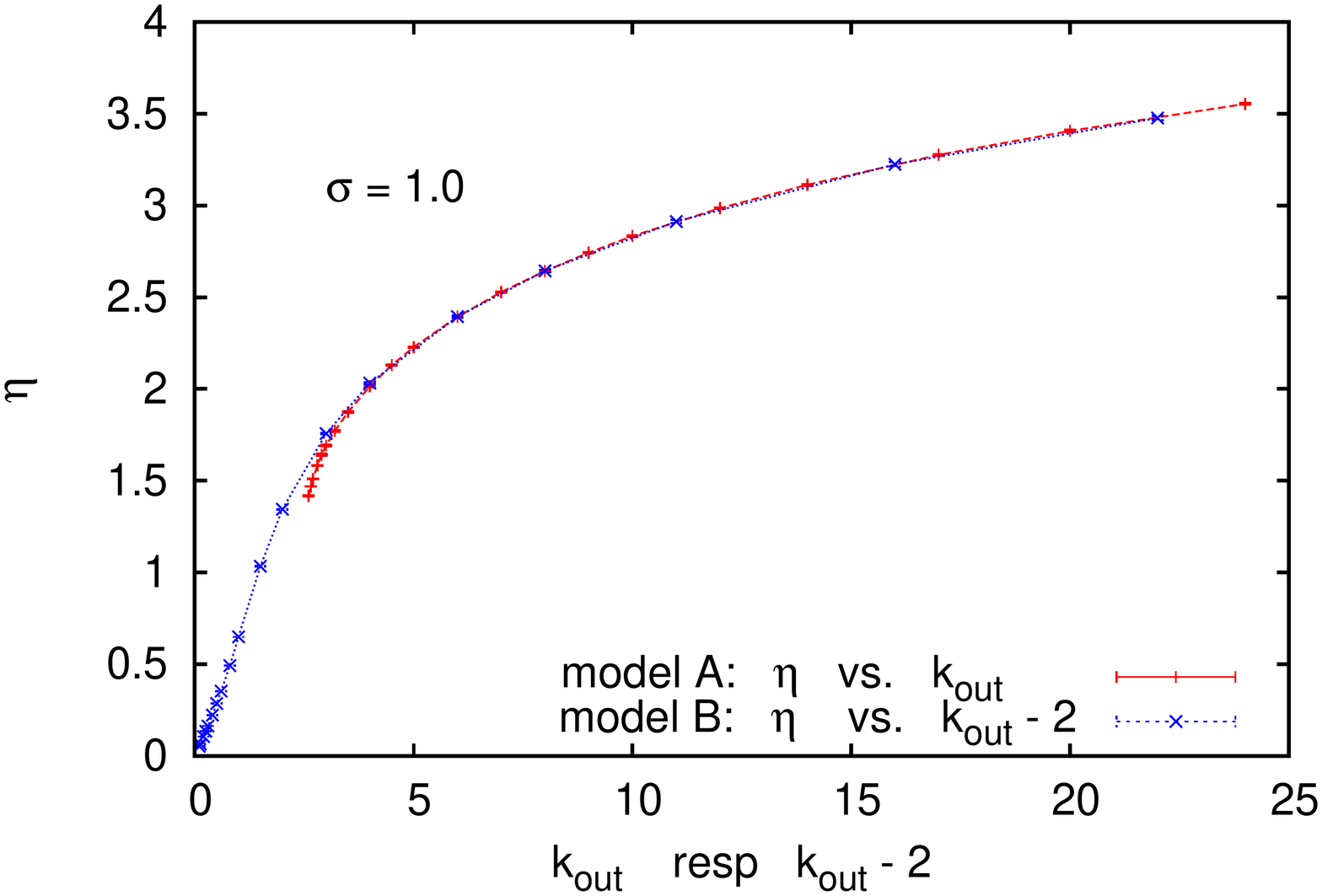}
\caption{(Color online) Exponents $\eta(k_{\rm out},\sigma=1)$ governing the increase of active 
sites obtained from Figs.~\ref{n_t-s20} and \ref{n_t-s20-sw}.  For both models, $\eta$ is plotted
against the number of {\it long-range} outgoing links, i.e. it is plotted against $k_{\rm out}$
for model (A) and against $k_{\rm out}-2$ for model (B).}
   \label{expon-n_t-s20}
\end{figure}

Numerical results for the increase of $n(t)$ in both models at $\sigma = 1$ are shown in 
Figs.~\ref{n_t-s20} to \ref{expon-n_t-s20}. From Fig.~\ref{n_t-s20} we see that model (A) 
exhibits indeed generic power law behavior (in agreement with the predictions of 
\cite{Anderson,Thouless,Cardy}) for all $k_{\rm out}>k_c$, where $k_c \approx 2.6(1)$:
\be
    n(t,k_{\rm out}) \sim t^{\eta(k_{\rm out})}\;.       \label{nuk}
\ee
At the critical point, a straightforward fit gives $\eta\equiv \eta(k_{\rm out}) = 1.42(4)$.

More precise determinations of $k_c$ and $\eta$ are possible by using the finite (cluster-)size 
scaling expected for BKT transitions. Let us define $\varepsilon = k_{\rm out}-k_c$. Near the 
critical point there exists a characteristic time scale $\tau(\varepsilon)$ which diverges as 
$\varepsilon \to 0$. Inversely, we can for each $t$ define an effective distance from the critical 
point as $\epsilon(t)$.
Using the latter we can make the finite-time scaling (FTS) ansatz

\be
   n(t,k_{\rm out}) \sim t^\eta g[\varepsilon/\epsilon(t))],   \label{nut-scaling}
\ee
where $g(z)$ is an analytic function joining smoothly the sub- and supercritical regions.
While this ansatz is common to models with short and long range infections, the scaling of 
$\epsilon(t)$ in the limit $t\to\infty$ is different. For short range contacts 
it is a power law, while for a BKT transition we expect
\be
   \epsilon(t) \sim [\ln \frac{t}{t_0}]^{-y}.  \label{tau-scaling}
\ee
Equations~(\ref{nut-scaling}) and (\ref{tau-scaling}) can be combined to 
\be
   n(t,k_{\rm out}) \sim t^\eta g[(k_{\rm out}-k_c) (\ln \frac{t}{t_0})^y]\;.
                    \label{nut-log-scaling}
\ee
A data collapse based on Eq.~(\ref{nut-log-scaling}) is shown in Fig.~\ref{fig-BKT-fit}. We see 
a nearly perfect collapse (only points with $t\geq 3$ are plotted), giving our best estimates
\be
   k_c = 2.60(1),\; \eta = 1.42(1),\; t_0=0.9(1) \quad {\rm and} \;\; y = 1.37(3). 
\ee
To our knowledge, neither $\eta$ nor $y$ have been calculated before. We conjecture that they
are universal for all models where $P(x)\sim 1/x^2$ asymptotically and where, in contrast to 
model (B), the epidemic can die (we made also preliminary simulations of a generalization of 
model (B) where left and right neighbors are infected with probabilities 0.9. The results 
support the conjecture). Our value of $k_c$ is consistent with the exact bound $k_c\geq 
2$ \cite{Aizenman} for this class of models.

Combining Eqs.~(\ref{nuk}) and (\ref{nut-log-scaling}) gives that $g(z)$ behaves for large $z$
as a stretched exponential, $\ln g(z) \sim z^{1/y}$, and that 
\be
   \eta(k_{\rm out}) -\eta \sim (k_{\rm out}-k_c)^{1/y}\;.
\ee
This is reasonably well satisfied.

\begin{figure}
\includegraphics[width=0.55\textwidth]{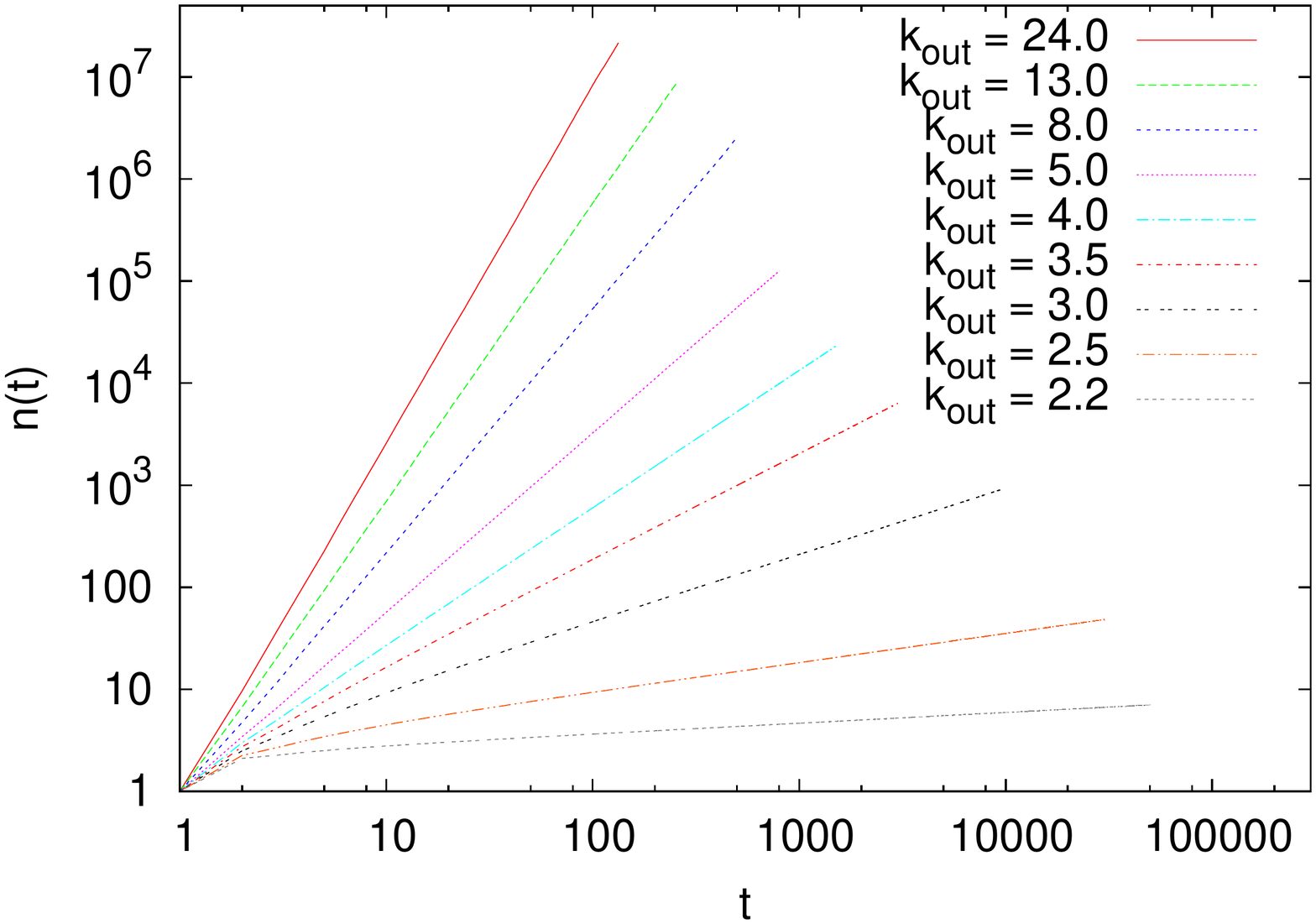}
\caption{(Color online) Same as Fig.~\ref{n_t-s20}, but for model (B). Since model (B) is 
supercritical for all $k_{\rm out}>2$, all curves now become straight lines for $t\to\infty$.}
   \label{n_t-s20-sw}
\end{figure}

As seen from Fig.~\ref{n_t-s20-sw}, essentially the same behavior is found also for model (B), 
with one important exception: Since model (B) with $\sigma \geq 1$ is supercritical for 
all $k_{\rm out}>2$, all curves become straight lines for $t\to\infty$ and $\eta(k_{\rm out})$
tends to zero for $k_{\rm out} \to 2$. The dependence of $\eta(k_{\rm out})$ on $k_{\rm out}$ 
is shown in Fig.~\ref{expon-n_t-s20} and Table 1. The fact that $n(t,k_{\rm out})$ increases as a 
power of $t$ is indeed known \cite{Coppersmith,Juhasz-2012}. In \cite{Coppersmith} exact upper bounds 
on $\eta(k_{\rm out})$ were given for small $k_{\rm out}$, and these were compared to simulation
results in \cite{Juhasz-2012}. When comparing our estimates with the results of \cite{Juhasz-2012},
we should notice that $1+\eta$ is the graph dimension of the cluster of immunes, and that the
constant $\beta$ used in \cite{Juhasz-2012} corresponds to $(k_{\rm out}-2)/2$. From 
Table 1 we see that our data are roughly 10 times more precise, but otherwise they are in very 
good agreement.

\begin{table}
\caption{Exponents $\eta(k_{\rm out})$ for model (B). Column 2  gives our results, column 3 is 
   from Table 1 of \cite{Juhasz-2012}, noticing that $d_g = 1+\eta$ and $\beta = (k_{\rm out}-2)/2$.}
\begin{tabular}{r|l|l}
\hline
\hline
\hspace*{0.4cm} $k_{\rm out} \qquad$  & $\qquad \eta$ (this work) $\qquad$ & $\qquad$ $\eta$ (Ref.~\cite{Juhasz-2012}) \hspace*{0.4cm}   \\
\hline
  \hspace*{0.4cm}  2.1  $\qquad$  & $ \qquad$ 0.0513(2)  $\qquad$  & $\qquad$                     \\
  \hspace*{0.4cm}  2.2  $\qquad$  & $ \qquad$ 0.1048(3)  $\qquad$  & $\qquad$      0.1038(24)      \\
  \hspace*{0.4cm}  2.3  $\qquad$  & $ \qquad$ 0.1625(4)  $\qquad$  & $\qquad$                      \\
  \hspace*{0.4cm}  2.4  $\qquad$  & $ \qquad$ 0.2226(4)  $\qquad$  & $\qquad$      0.2121(44)      \\
  \hspace*{0.4cm}  2.5  $\qquad$  & $ \qquad$ 0.2855(6)  $\qquad$  & $\qquad$                      \\
  \hspace*{0.4cm}  2.6  $\qquad$  & $ \qquad$ 0.3524(7)  $\qquad$  & $\qquad$      0.3532(74)      \\
  \hspace*{0.4cm}  2.8  $\qquad$  & $ \qquad$ 0.4963(7)  $\qquad$  & $\qquad$      0.4992(67)      \\
  \hspace*{0.4cm}  3.0  $\qquad$  & $ \qquad$ 0.6492(6)  $\qquad$  & $\qquad$       0.656(8)       \\
  \hspace*{0.4cm}  3.5  $\qquad$  & $ \qquad$ 1.0330(17) $\qquad$  & $\qquad$                      \\
  \hspace*{0.4cm}  4.0  $\qquad$  & $ \qquad$ 1.344(2)   $\qquad$  & $\qquad$       1.347(16)      \\
  \hspace*{0.4cm}  5.0  $\qquad$  & $ \qquad$ 1.757(4)   $\qquad$  & $\qquad$       1.770(23)      \\
  \hspace*{0.4cm}  6.0  $\qquad$  & $ \qquad$ 2.032(4)   $\qquad$  & $\qquad$                      \\
  \hspace*{0.4cm}  8.0  $\qquad$  & $ \qquad$ 2.394(6)   $\qquad$  & $\qquad$                      \\
  \hspace*{0.4cm} 10.0  $\qquad$  & $ \qquad$ 2.646(7)   $\qquad$  & $\qquad$                      \\
  \hspace*{0.4cm} 13.0  $\qquad$  & $ \qquad$ 2.913(9)   $\qquad$  & $\qquad$                      \\
  \hspace*{0.4cm} 18.0  $\qquad$  & $ \qquad$ 3.224(13)  $\qquad$  & $\qquad$                      \\
  \hspace*{0.4cm} 24.0  $\qquad$  & $ \qquad$ 3.478(12)  $\qquad$  & $\qquad$                      \\
\hline
\hline
\end{tabular}
\end{table}

For both models $\eta$ becomes the same for large $k_{\rm out}$, which is to be expected: 
For large $k_{\rm out}$, the evolution mainly depends on long range infections, as 
infections between nearest neighbors just fill in gaps in the cluster of immunes. It 
seems that $\eta \sim \log k_{\rm out}$ for large $k_{\rm out}$, so that 
$n(t) \sim \exp(\ln k_{\rm out} \ln t)$. More precisely, both data sets 
are very well fitted by $\eta = 0.9+0.84 \ln(k_{\rm out}-k_{\rm out, short})$,
where $k_{\rm out, short}$ is the number of short-range infections.  
We have no theoretical argument for this.

\begin{figure}
\includegraphics[width=0.55\textwidth]{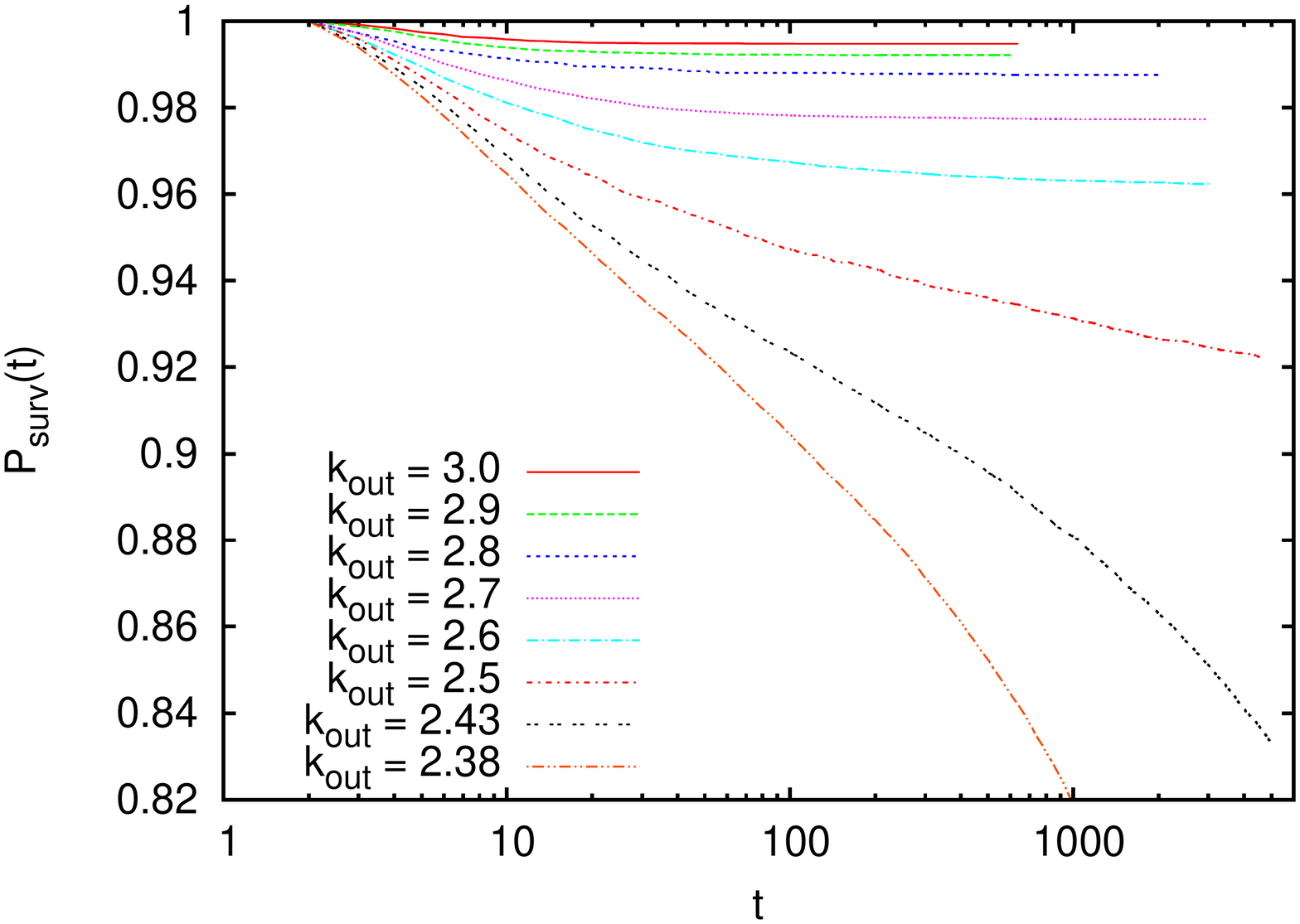}
\caption{(Color online) Log-linear plot of $P_{\rm surv}(t)$, the probability that there are 
still some active sites at time $t$, for model (A) with $\sigma=1$. All curves for  
$k_{\rm out}\geq k_c$ (including the critical one!) seem to become horizontal with 
$P_{\rm surv}(\infty)>0$.}
   \label{P_t-s20}
\end{figure}

Survival probabilities $P_{\rm surv}(t)$ for model (A) are shown in Fig.~\ref{P_t-s20}. 
As expected, they all tend to constants when the process is supercritical. More surprising,
it seems than also for the critical case $k_{\rm out}=k_c=2.6$ we have $P_{\rm surv}(t)\to const$
for $t\to\infty$. Indeed, even in the clearly subcritical case $k_{\rm out}=2.4$, we see that 
$P_{\rm surv}(t)$ first curves upward, before it finally goes to zero. This is in contrast
to the behavior of SIR epidemics with short range contacts, but it is precisely what was 
proven rigorously in \cite{Aizenman}: When $\sigma=1$, the percolation transition is 
discontinuous in the sense that the order parameter (which is just $\lim_{t\to \infty} P_{\rm surv}(t)$)
jumps discontinuously when the control parameter $k_{\rm out}$ passes through the critical 
point. Furthermore, $P_{\rm surv}(\infty) > 2/k_c$ as predicted in \cite{Aizenman}.

\begin{figure}
\includegraphics[width=0.55\textwidth]{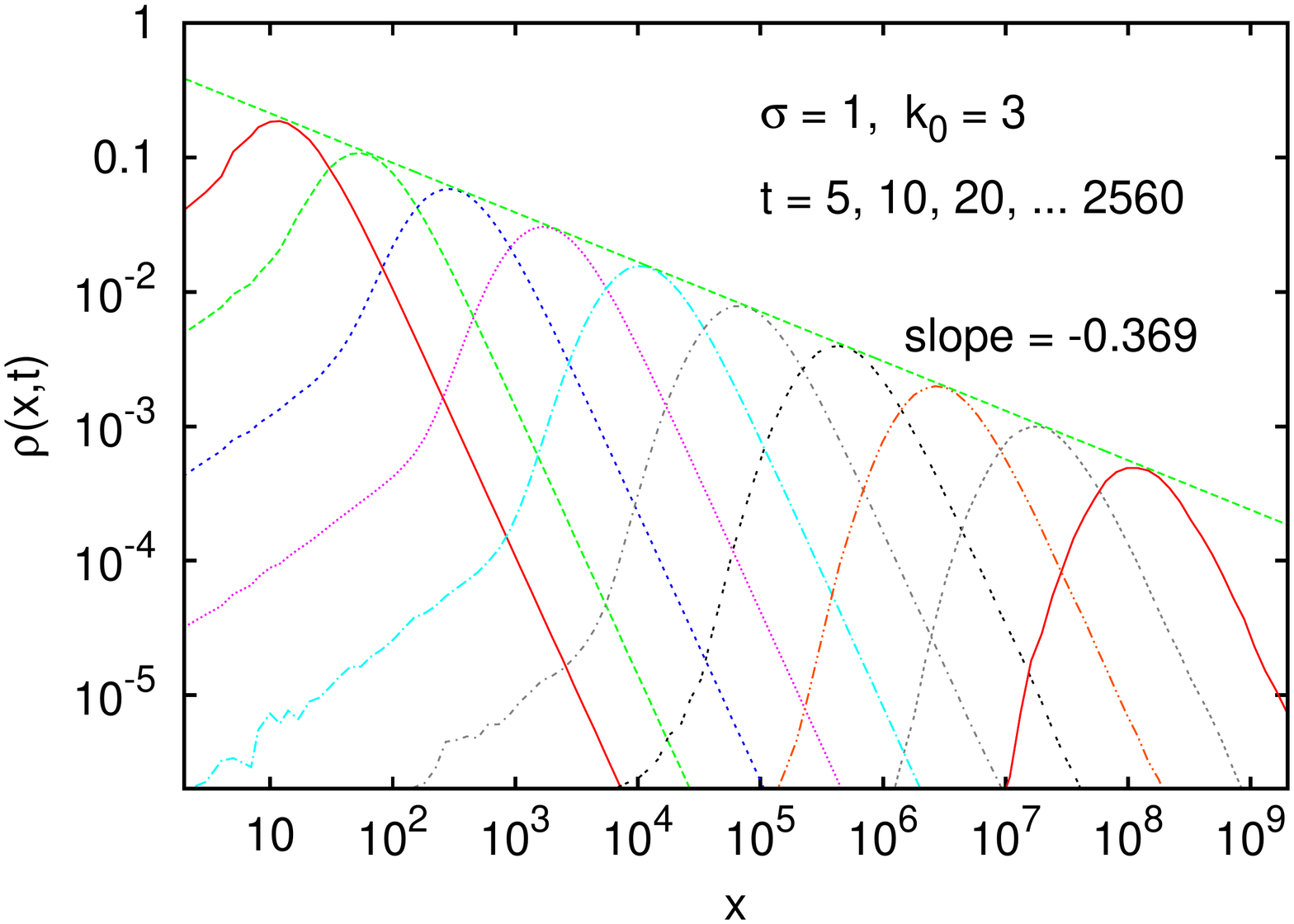}
\caption{(Color online) Log-log plot of the densities of active sites for model (A) with 
$\sigma=1$ and $k_{\rm out}=3$, for times $t=5,10,\ldots 2560$.}
  \label{act_t-s20}
\end{figure}

\begin{figure}
\includegraphics[width=0.55\textwidth]{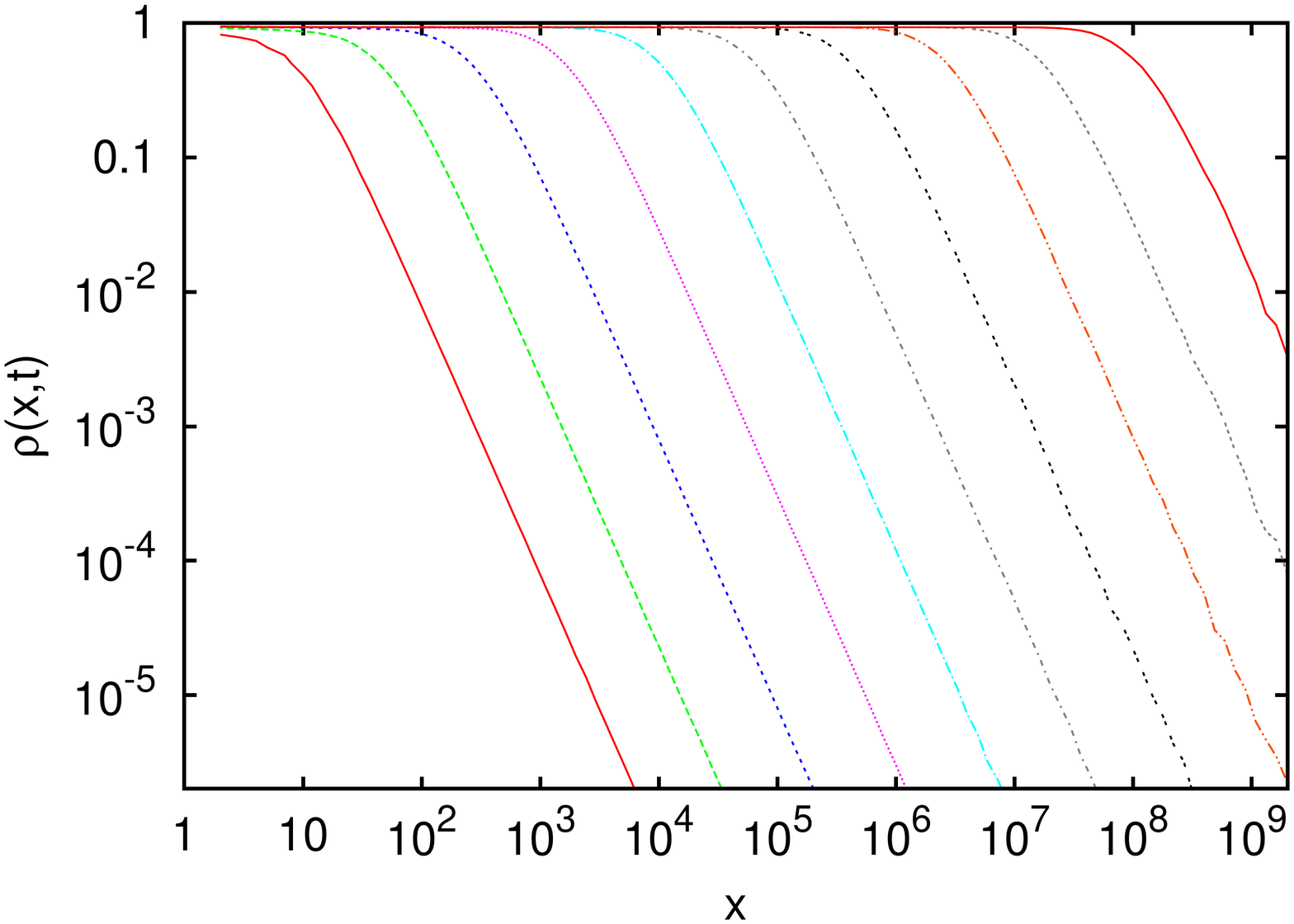}
\caption{(Color online) Log-log plot of the densities of immune sites for model (A) with 
the same parameters as in Fig.~\ref{act_t-s20}.}
  \label{all_t-s20}
\end{figure}

Spatial distributions of active and immune sites for model (A) are shown in Figs.~\ref{act_t-s20}
and \ref{all_t-s20}.
Superficially, these distributions look similar to those for the supercritical case with
$\sigma <1$ shown in Figs.~\ref{active-s15} and \ref{immune-s15}, but there are important differences:

\begin{itemize}
\item The curves are now equidistant for exponentially increasing times, corresponding 
to the fact that the number of active sites, the number of immune sites, and the effective
radius all increase like powers of $t$, while they increased faster than polynomial
when $\sigma <1$.
\item While the peaks in Fig.~\ref{active-s15} become narrower with increasing $t$, now the 
shapes of the curves are independent of $t$, suggesting finite time scaling for 
$\sigma =1$, but not for $\sigma <1$. This is also seen by making formal data collapses
(not shown here).

In view of the fact that $R(x,t)$ is constant ($t$-independent) for small values of $x$ 
we make for $\sigma =1$ and $k_{\rm out}\geq k_c$ the ansatz
\be
   R(x,t) \approx \phi(x/\xi(t))           \label{Rphi}
\ee
for $t\to\infty$, with
\be
    \phi(z) = \left\{
            \begin{array}{rl}
            \text{ const } & \text{ for } |z| \ll 1,\\
            |z|^{-\sigma-1}  & \text{ for } |z| \gg 1 .
            \end{array} \right.
\label{scaling}
\ee
and $\xi(t)\to\infty$ for $t\to\infty$. From this ansatz follows immediately that $\xi(t)$
has to scale exactly like $N(t)$,
\be
    \xi(t)\sim t^{1+\eta},                  \label{xi}
\ee
and that the density of active sites satisfies the scaling
\be
  \rho(x,t) \approx t^{-1} \psi(x/\xi(t))     \label{rpsi}
\ee
with $\psi(z)=z \phi'(z).$

\item Since, according to Eqs.(\ref{rpsi}) and (\ref{xi}), both the peak heights and the 
peak positions of $\rho(x,t)$ scale as powers of $t$ for $\sigma=1$, also the peak heights 
must scale as a power of the positions,
\be
   \rho_{\rm max}(t) \equiv \max_x \rho(x,t) \sim \xi(t)^{-1/(1+\eta)}\;.   \label{eta-eta}
\ee
This is clearly seen in Fig.~\ref{act_t-s20}, where we obtain $1/(1+\eta) = 0.369(3)$ or 
$\eta=1.71(2)$, in perfect agreement with the direct measurement $\eta=1.692(6)$. But as 
we shall see later, Eq.~(\ref{eta-eta}) does not hold for $\sigma < 1$.
\end{itemize}

Since the generated clusters are basically compact, we can measure $\eta$ also by 
measuring the average of $\log |x|$, either over the active or over the immune sites.
Both averages should scale as $\log \xi(t) \approx (1+\eta) \log t$. This was indeed
verified numerically.

Equations (\ref{Rphi}) to (\ref{eta-eta}) were also tested for several other values of 
$k_{\rm out}$, and were satisfied in all cases. This includes even the critical case $k_c$.
Also there, $R(x,t)$ is flat for $x\ll \xi(t)$, showing that the pair connectedness does
not decrease as long as $x<\xi(t)$. This is of course in agreement with the result that 
$P_{\rm surv}$ does not decrease to zero at the critical point, and that the percolation
transition is discontinuous for $\sigma=1$ \cite{Aizenman}.

\begin{figure}
\includegraphics[width=0.55\textwidth]{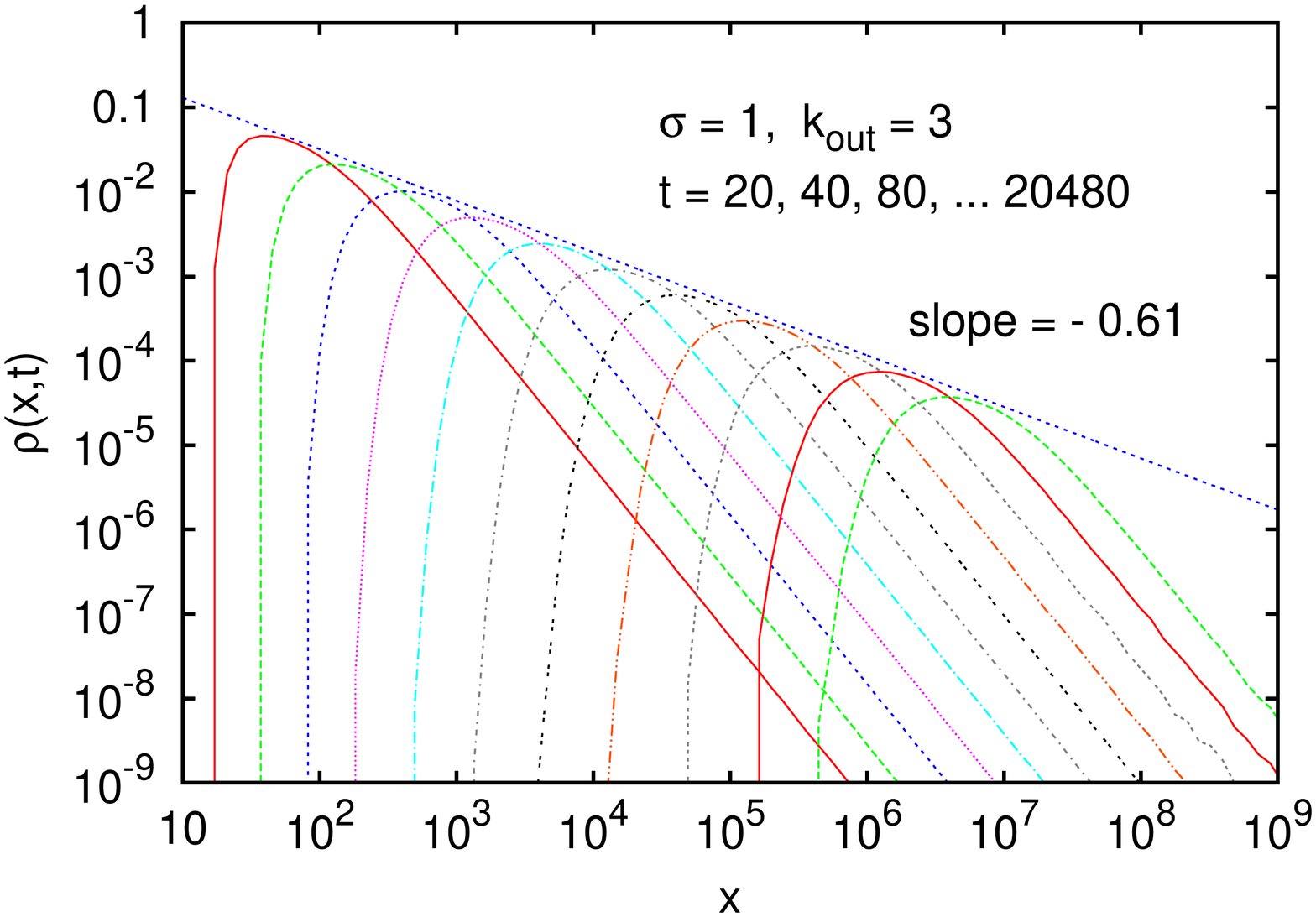}
\caption{(Color online) Log-log plot of the densities of active sites for model (B) with 
$\sigma=1$ and $k_{\rm out}=3$, for times $t=20,40,\ldots 20480$. Compared to 
Fig.~\ref{act_t-s20}, the curves for the smallest $t$ values now show less perfect 
scaling.}
  \label{act_sw-t-s20}
\end{figure}

We should point out that Eqs.~(\ref{Rphi}) to (\ref{eta-eta}) hold also for 
model (B) if $\sigma=1$, although finite time corrections are somewhat larger since
not all links are distributed according to a power law. For one typical case 
$(k_{\rm out} = 3.0)$, see Fig.~\ref{act_sw-t-s20}. In spite of the visible deviation
in the curve for $t=20$, the peaks line up for larger $t$ along a perfect power law,
with $\eta = 1-1/0.610(4) = 0.64(1)$. This should be compared to $\eta = 0.6484(4)$ 
from the direct measurement of $n(t)$. 

Finally, let us discuss the predictions of mean field theory for the case $\sigma=1$. 
The exact evolution for $R(x,t)$ can be written as 
\bea
   R(x,t+1) & = & R(x,t)                 \\
        & + & \sum_y P(x-y) \langle n_{\rm act}(y,t) (1-n_{\rm immune}(x,t)) \rangle\;, \nonumber
                          \label{mf-sigma2}
\eea
where $n_{\rm act}(x,t)$ and $n_{\rm immune}(x,t)$ are the exact fluctuating densities
of active and immune sites, and angular brackets indicate an ensemble average. The mean 
field assumption is 
\bea
   \langle n_{\rm act}(y,t) n_{\rm immune}(x,t) \rangle & = &     \nonumber \\
               & = & \langle n_{\rm act}(y,t) \rangle \langle n_{\rm immune}(x,t) \rangle \nonumber \\
               & = & \rho(x,t) R(x,t).
\eea
Inserting this and the scaling ansatzes into the r.h.s. of Eq.(\ref{mf-sigma2}), we 
find that it gives back the scaling ansatz for the l.h.s., i.e. our scaling assumptions
are consistent with mean field theory. In contrast, the scaling ansatzes would not be 
compatible with this mean field theory, if $\sigma \neq 1$. 

\section{The critical model (A) for $0<\sigma < 1$}    \label{critical}

Critical percolation with long range infection has been studied both by means of 
field theory \cite{Janssen99,Linder} and by means of simulations \cite{Linder}.
The field theoretic analysis (using the epsilon expansion) should hold in $d$
dimension of space for any $d$, provided $\sigma \approx d/3$ \cite{Janssen99,Linder} -- 
thus it should also be applicable to $d=1$, provided $\sigma \approx 1/3$. Remarkably, 
it predicts that mean field theory does not only hold for $\sigma<0$ (where it holds
for the supercritical case), but also for $0< \sigma < 1/3$. It also predicts, for 
$d\geq 2$, the value of $\sigma$ above which short range behavior should be observed.
Again it is remarkable that this value is not where short range behavior is observed
in the supercritical case ($\sigma =d$), but at a smaller value of $\sigma$ \cite{Linder}.

While these predictions have been reasonably well confirmed for $d=2$ \cite{Linder} 
(much larger simulations, again on lattices with $2^{64}$ sites, will be published
elsewhere \cite{Grassberger2013}), it seems that no simulations were yet done 
for $d=1$.

\begin{figure}
\includegraphics[width=0.55\textwidth]{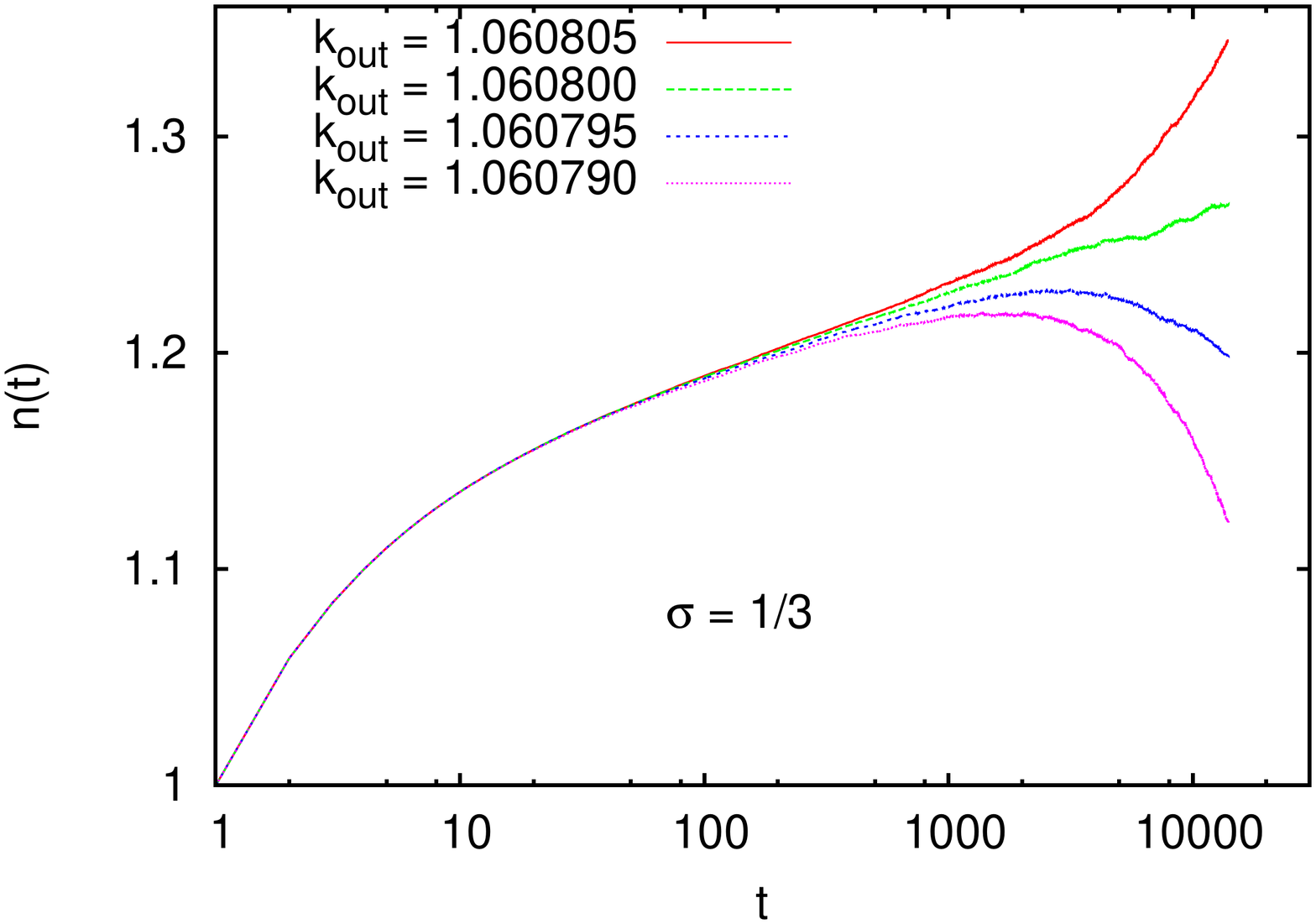}
\caption{(Color online) Log-linear plot of $n(t)$ for model (A) with
$\sigma=1/3$ and $k_{\rm out}\approx k_c$. These data are compatible with $k_c = 1.06799(1)$
and $\eta=0$, if we allow for logarithmic corrections, $n(t) \sim [\ln t]^\alpha$ with
$\alpha \approx 0.5$ to 1.}
  \label{fig-n_t-alpha3}
\end{figure}

\begin{figure}
\includegraphics[width=0.55\textwidth]{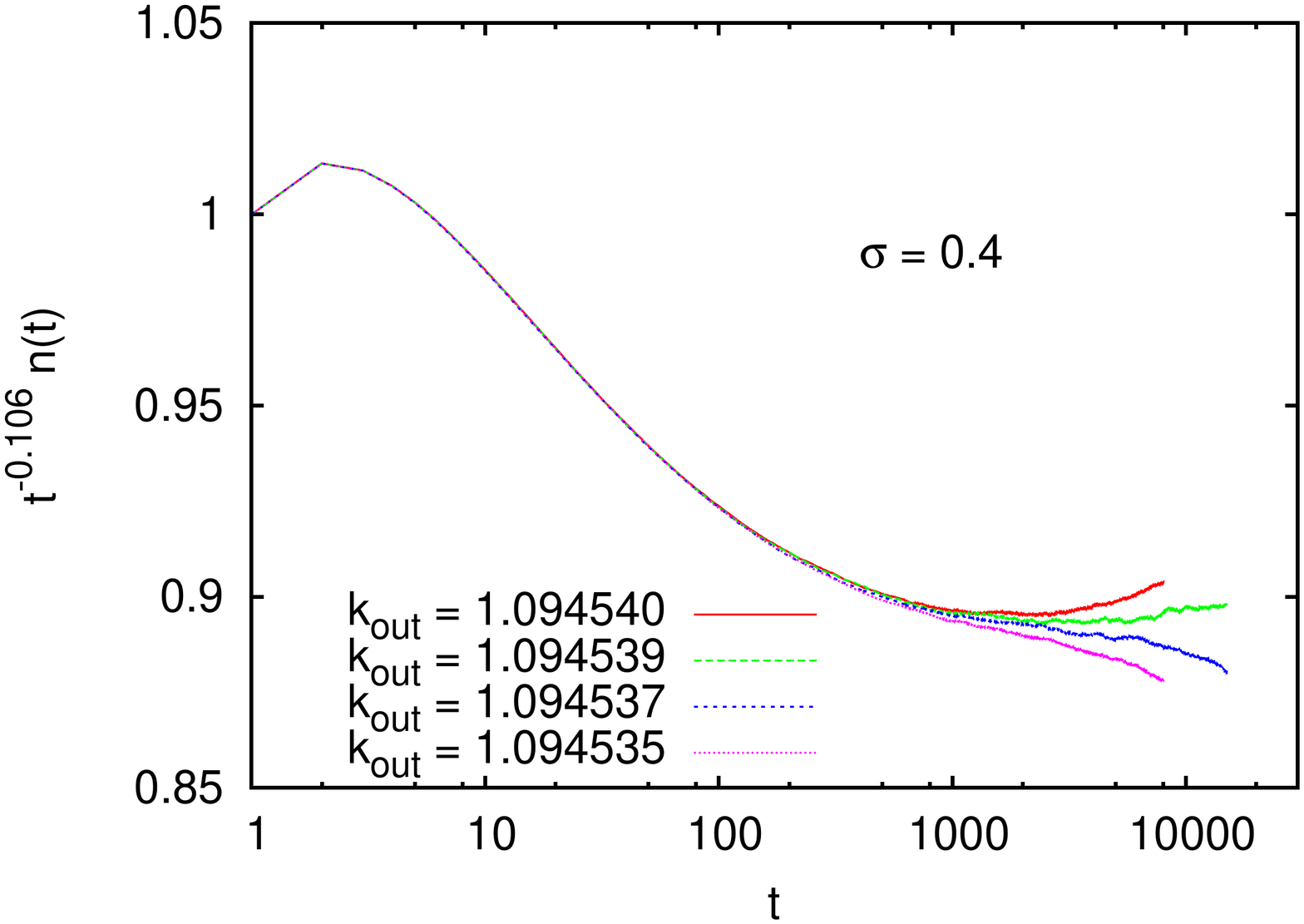}
\caption{(Color online) Log-linear plot of $t^{-0.106}n(t)$ for model (A) with
$\sigma=0.4$ and $k_{\rm out}\approx k_c$. These data are compatible with $k_c = 1.094538(2)$
and $\eta=0.105(3)$. The main uncertainty of $\eta$ comes from the obvious corrections to 
scaling. Notice, however, the expanded $y$-scale. Notice also that finite (lattice) size
corrections are negligible in these simulations, even for the largest values of $t$. Thus,
in order to obtain the true critical exponents, one should take the large-$t$ behavior 
serious, and should not try to fit in an intermediate region of $t$.
}
  \label{fig-n_t-alpha25}
\end{figure}

The most obvious strategy for finding the critical values $k_c$ of $k_{\rm out}$ is to 
start again from a single seed and to look for the best scaling behavior $n(t)\sim t^\eta$
in the large-$t$ limit. In general this works without too mGany problems, but we 
have to expect large finite-$t$ corrections near any change of the critical behavior,
i.e. in particular near $\sigma=1/3$ and near $\sigma=1$. But one must be aware of 
surprises. As two examples we show 
$n(t)$ versus $t$ for $\sigma = 1/3$ (where we expect $\eta=0$, but slow
convergence due to possible logarithmic corrections) and for $\sigma = 0.4$, where 
we should a priori expect much less problems. The data, based on simulations of 
typically $\approx 10^8$ clusters,  are shown in Figs.~\ref{fig-n_t-alpha3} and 
\ref{fig-n_t-alpha25}. While Fig.~\ref{fig-n_t-alpha3} is compatible with nothing 
more than the expected logarithmic corrections, the corrections to scaling seen in 
Fig.~\ref{fig-n_t-alpha25} are much more complicated. They cannot be described by a 
single power term, and they contribute most to the uncertainty of our estimate of 
$\eta$. 

Plots similar to Figs.~\ref{fig-n_t-alpha3} and \ref{fig-n_t-alpha25} were also made 
for several other values of $\sigma$. In addition to $n(t)$, we also looked in the 
same way at $P_{\rm surv}(t)$ (plots not shown), for which we assumed
\be
   P_{\rm surv}(t) \sim t^{-\delta}.
\ee

\begin{figure}
\includegraphics[width=0.55\textwidth]{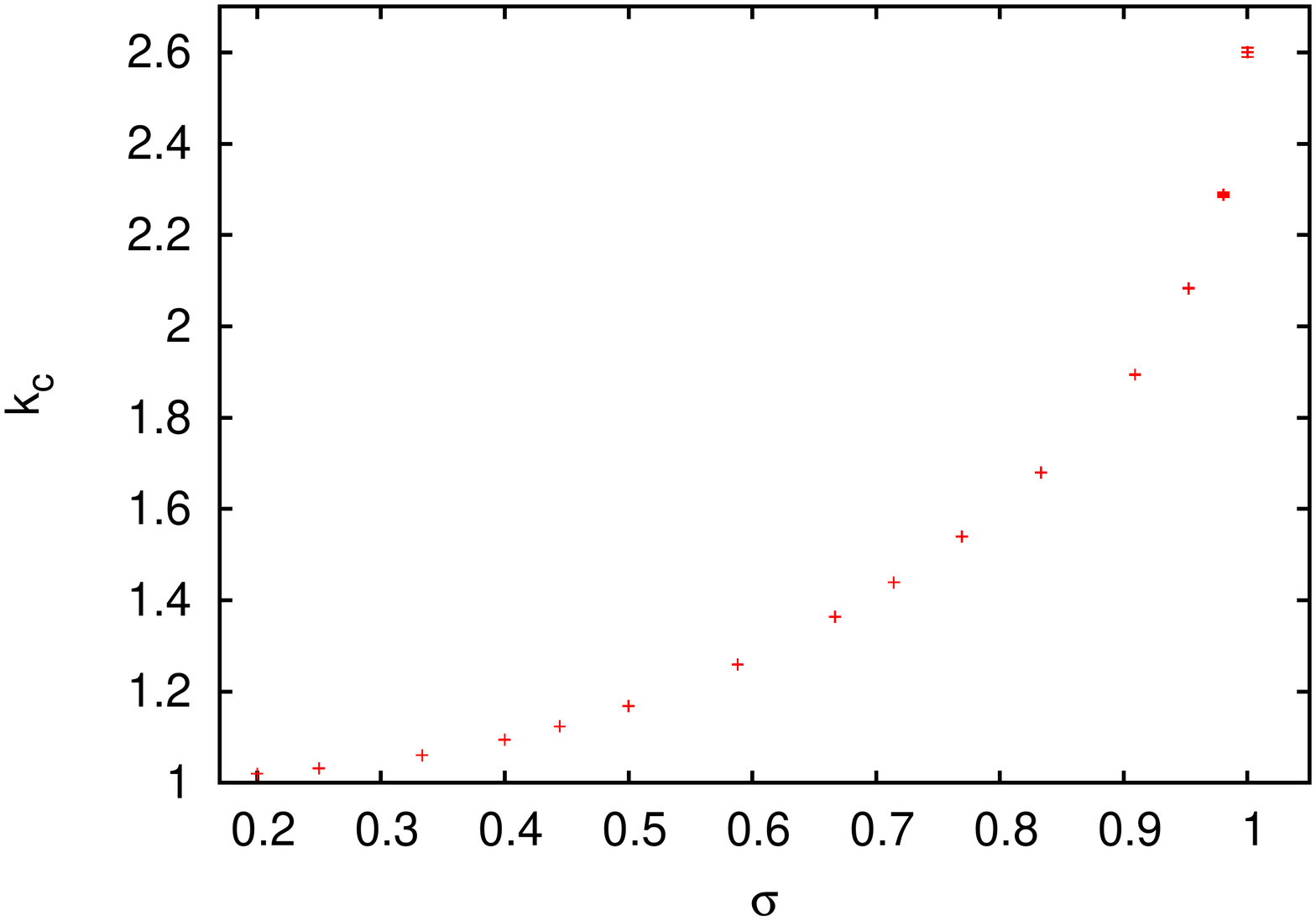}
\caption{(Color online) Values of $k_c$ versus $\sigma$
for critical model (A). For small $\sigma$ the values approach the value $k_c \to 1$ for 
branching processes. For $\sigma\to 1$ the value $k_c=2.60(1)$ obtained in the last section 
is reached with an infinite slope.} 
  \label{fig-k_c}
\end{figure}

\begin{figure}
\includegraphics[width=0.55\textwidth]{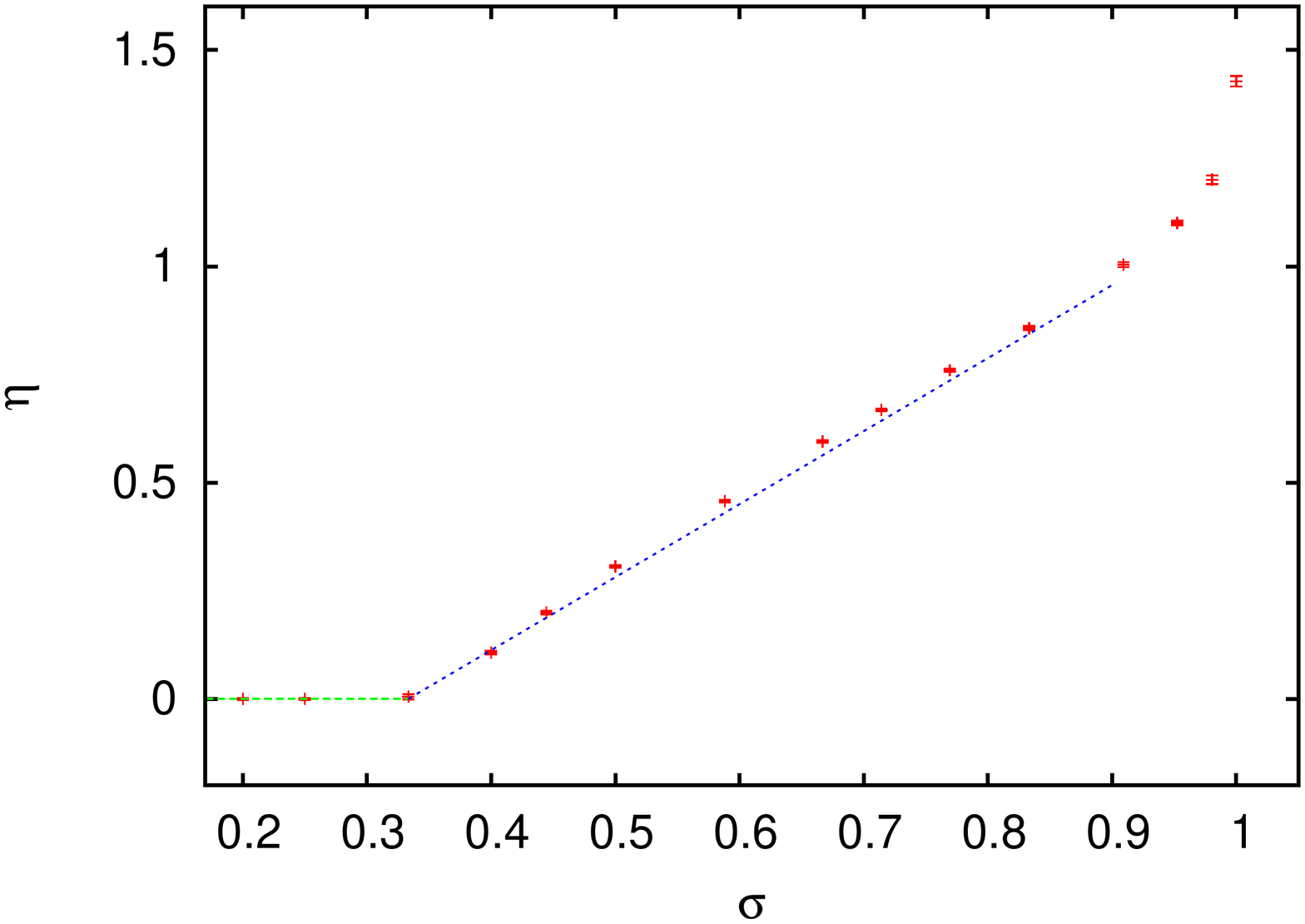}
\caption{(Color online) Exponents $\eta(\sigma)$, defined via $n(t)\sim t^\eta$, versus $\sigma$
for critical model (A). The horizontal line is the mean field prediction $\eta=0$ for $\sigma<1/3$. 
The straight tilted line is the prediction of \cite{Janssen99} that should be exact for 
$\sigma$ slightly above $1/3$. Finally for $\sigma=1$ we have the value $\eta=1.428(12)$ 
obtained in Sec.~III.}
  \label{fig-eta}
\end{figure}

\begin{figure}
\includegraphics[width=0.55\textwidth]{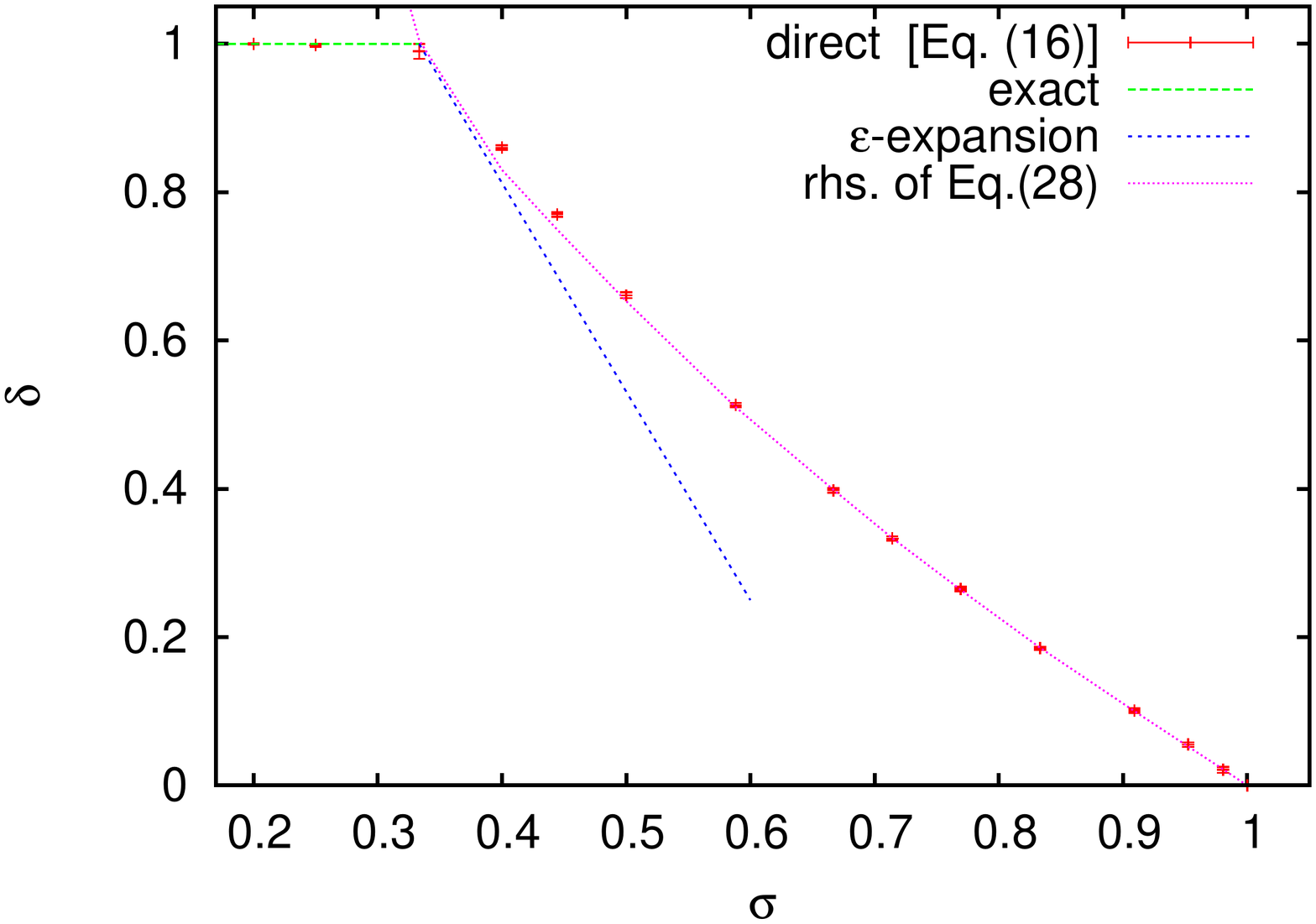}
\caption{(Color online) Exponent $\delta$ versus $\sigma$ for critical model (A). 
The points with error bars are obtained directly from the definition $P_{\rm surv}(t)\sim t^{-\delta}$, 
while the dotted curve is obtained from the data shown in Fig.~\ref{fig-eta}, using Eq.~(\ref{delta-eta}).
Again the straight lines are the exact mean field prediction
(for $\sigma<1/3$) and the epsilon expansion result of \cite{Janssen99} (for $\sigma$ slightly
above 1/3).}
  \label{fig-delta}
\end{figure}

Results are shown in Figs.~\ref{fig-k_c} to \ref{fig-delta}. They first of all confirm the 
prediction that mean field behavior holds for all $\sigma < 1/3$, not only for $\sigma < 0$ 
as in the supercritical case. The critical values of $k_{\rm out}$ converge to $k_c\to 1$ 
for $\sigma \to 0$. Both $k_c$ and $\eta$ seem to reach their limits for $\sigma \to 1$ with 
infinite slope, $\lim_{\sigma \to 1}dk_c/d\sigma = \lim_{\sigma \to 1}d\eta/d\sigma = \infty$. 
This is to be expected, since both can be viewed as infinite for $\sigma >1$: There, the 
epidemic dies out for any finite value of $k$ -- but as $k$ is increased, the mass increases 
during the transient faster than any power. We also see, in agreement with the last section, 
that $\delta \to 0$ when $\sigma \to 1$. On the basis of 
Fig.~\ref{fig-delta} we conjecture more precisely that $\delta \approx 1-\sigma$ when 
$\sigma \to 1$. Finally we compare to the predictions of \cite{Janssen99}, represented in 
Figs.~\ref{fig-eta} and \ref{fig-delta} by straight lines. These predictions are based 
on an $\epsilon$-expansion with $\epsilon = d-3\sigma$. Although this expansion is 
most likely only asymptotic, it should give correct results when $\epsilon \to 0$. 
This is definitely the case for Fig.~\ref{fig-eta}, where the prediction seems to be 
valid up to $\sigma\approx 1$. 
For $\delta$ the agreement is much worse. Even for $\sigma = 0.4$
it gives a prediction for $1-\delta$ that is too large by $\approx 15$ per cent. This 
might be related to the anomaly seen in Fig.~\ref{fig-n_t-alpha25}, and might 
indicate that we have still underestimated finite-$t$ corrections in the regime
$1/3<\sigma < 1/2$. The latter is also suggested by the slight disagreement in the
region $1/3 < \sigma < 1/2$ with Eq.~(\ref{delta-eta}).

\begin{figure}
\includegraphics[width=0.55\textwidth]{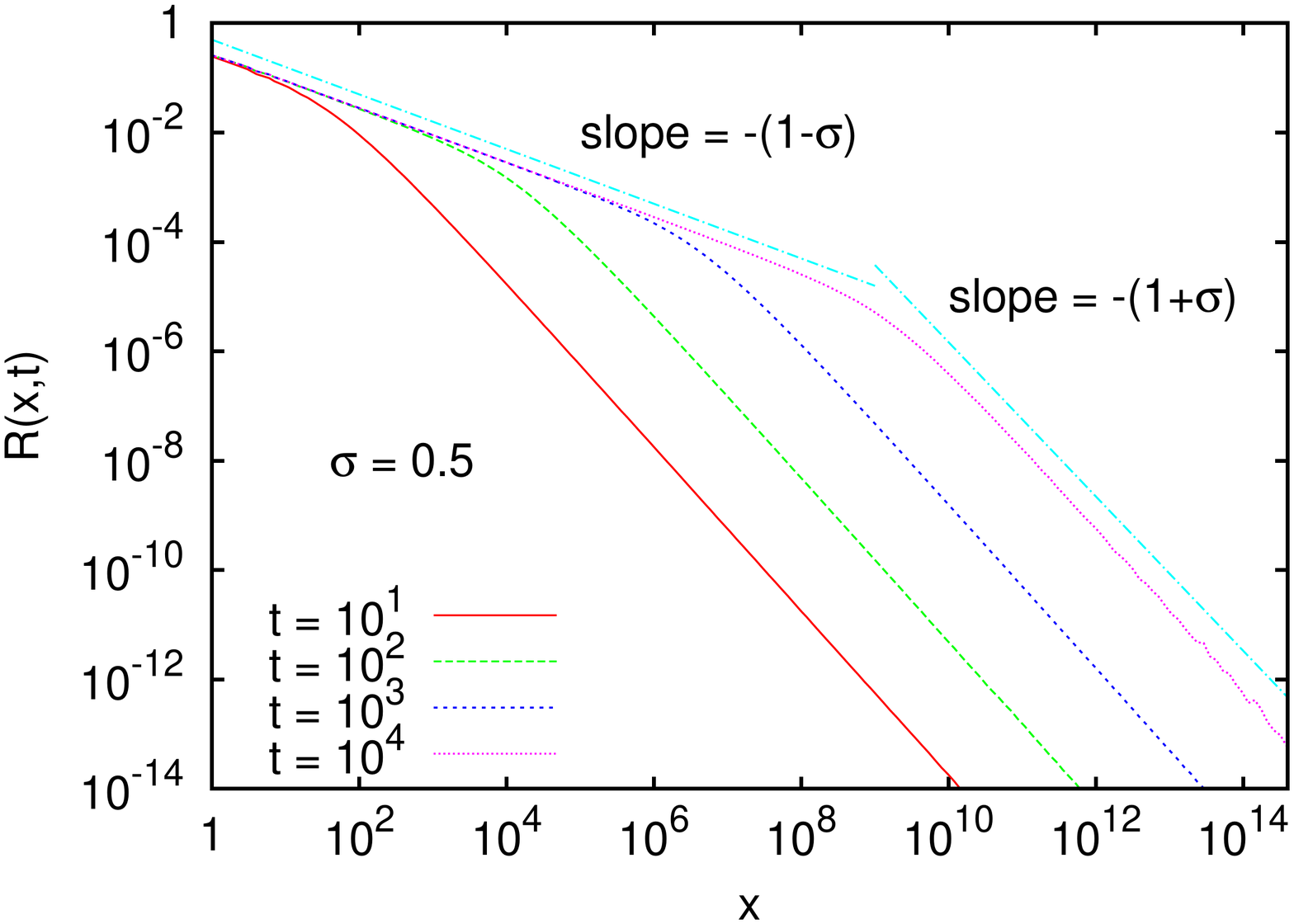}
\caption{(Color online) Densities of immune sites for critical epidemics with $\sigma=1/2$, 
for $t = 10, 10^2, \ldots 10^4$. For large $x$ all curves show the by now well known 
tail $R\propto 1/x^{1+\sigma}$, but for smaller values of $x$ the flat part seen in 
Figs.~3, 4 and 11 is replaced by a power law $R\propto 1/x^{1-\sigma}$.}
  \label{fig-all-crit}
\end{figure}

For SIR epidemics with short range infection, one can also define a critical exponent
$z$ that describes how the correlation length increases with time, $\xi\sim t^z$.
Figure \ref{fig-all-crit} suggests that there is indeed a well defined diverging length
scale (the location where the break of slopes occurs; notice that this length scale
should be defined by geometric averaging \cite{Grassberger86}, instead of the arithmetic
averaging usually taken for short range infects -- characterizing length scales by arithmetic
averages when $\sigma <1$, as done e.g. in \cite{Emmerich}, can lead to dubious
results). In that figure we present the densities of 
immune sites for different values of $t$, for one randomly picked value of $\sigma$.
The data shown in Fig.~\ref{fig-all-crit} are for $\sigma=1/2$, but similar results were 
found for all other $0 < \sigma < 1$. In particular, in all cases we see two different
powers for $x<\xi(t)$ and $x> \xi(t)$.

More precisely, we observe the by now well known tail $R(x,t) \propto 1/x^{1+\sigma}$
for large $x$, but for small $x$ we observe a completely new phenomenon. Instead of 
the flat parts seen in Figs.~3, 4 and 11 for small $x$ we now see another power 
law, 
\be
   R(x,t) \propto 1/x^{1-\sigma},
\ee
for $x<\xi(t)$, where $\xi(t)$ is a function that increases like a power for large $t$,
\be
    \xi(t) \sim t^z.                  \label{xit}
\ee
Indeed, data collapses (not presented here) show that $R(x,t)$ satisfies a scaling law 
similar to Eq.~\ref{Rphi},
\be
   R(x,t) = x^{\sigma-1} \Phi(x/\xi(t))\;,   \label{RPhi}
\ee
where 
\be
   \Phi(x) = \left\{
            \begin{array}{rl}
            \text{ const } & \text{ for } |x| \ll 1,\\
            |x|^{-2\sigma}  & \text{ for } |x| \gg 1 .
            \end{array} \right.
     \label{crit-scaling}
\ee

It is not difficult to relate the exponent $z$ to the exponent 
$\eta$. Indeed, by summing $R(x,t)$ over all $x$ we obtain
\bea
   N(t) &   =     & \sum_x R(x,t)  \nonumber \\
        & \approx & \int_0^{\xi(t)} \frac{dx}{x^{1-\sigma}} + [\xi(t)]^{2\sigma} 
                                        \int_{\xi(t)}^\infty \frac{dx}{x^{1+\sigma}} \nonumber \\
        &   =     & \frac{[\xi(t)]^\sigma}{\sigma} + \frac{[\xi(t)]^\sigma}{\sigma}  \nonumber \\
        &   =     & \frac{2[\xi(t)]^\sigma}{\sigma}\;,
\eea
where we used that $0<\sigma < 1$. Since $N(t)\sim t^{1+\eta}$, we have thus \cite{Janssen99}
\be
    z = \frac{1+\eta}{\sigma}.     \label{z-eta}
\ee

For $\sigma \to 1$ this gives $\xi(t) \sim t^{1+\eta}$ as found in Sec.~III, while 
it gives $z\to \infty$ for $\sigma\to 0$, indicating that in this limit $\xi(t)$ is infinite 
and $R(x,t)\sim 1/t$ is a pure power law for all $t$.

\begin{figure}
\includegraphics[width=0.55\textwidth]{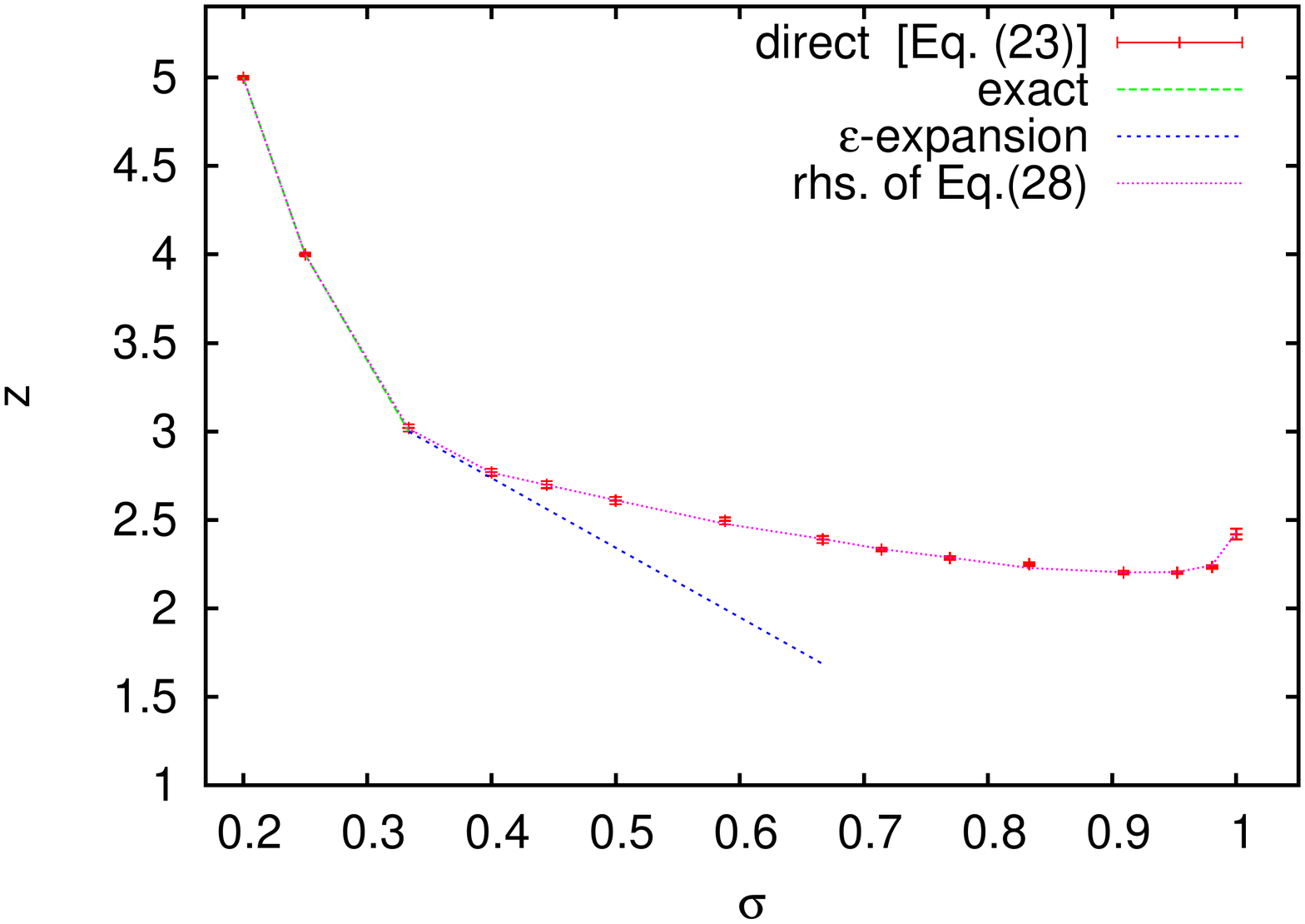}
\caption{(Color online) Critical exponent $z$ versus $\sigma$. As in Fig.~\ref{fig-delta}, 
points with error bars are from direct measurements (using now Eq.~(\ref{defX})), while the dotted curve 
is the prediction from Eq.(\ref{z-eta}).}
  \label{fig-z-crit}
\end{figure}

\begin{figure}
\includegraphics[width=0.55\textwidth]{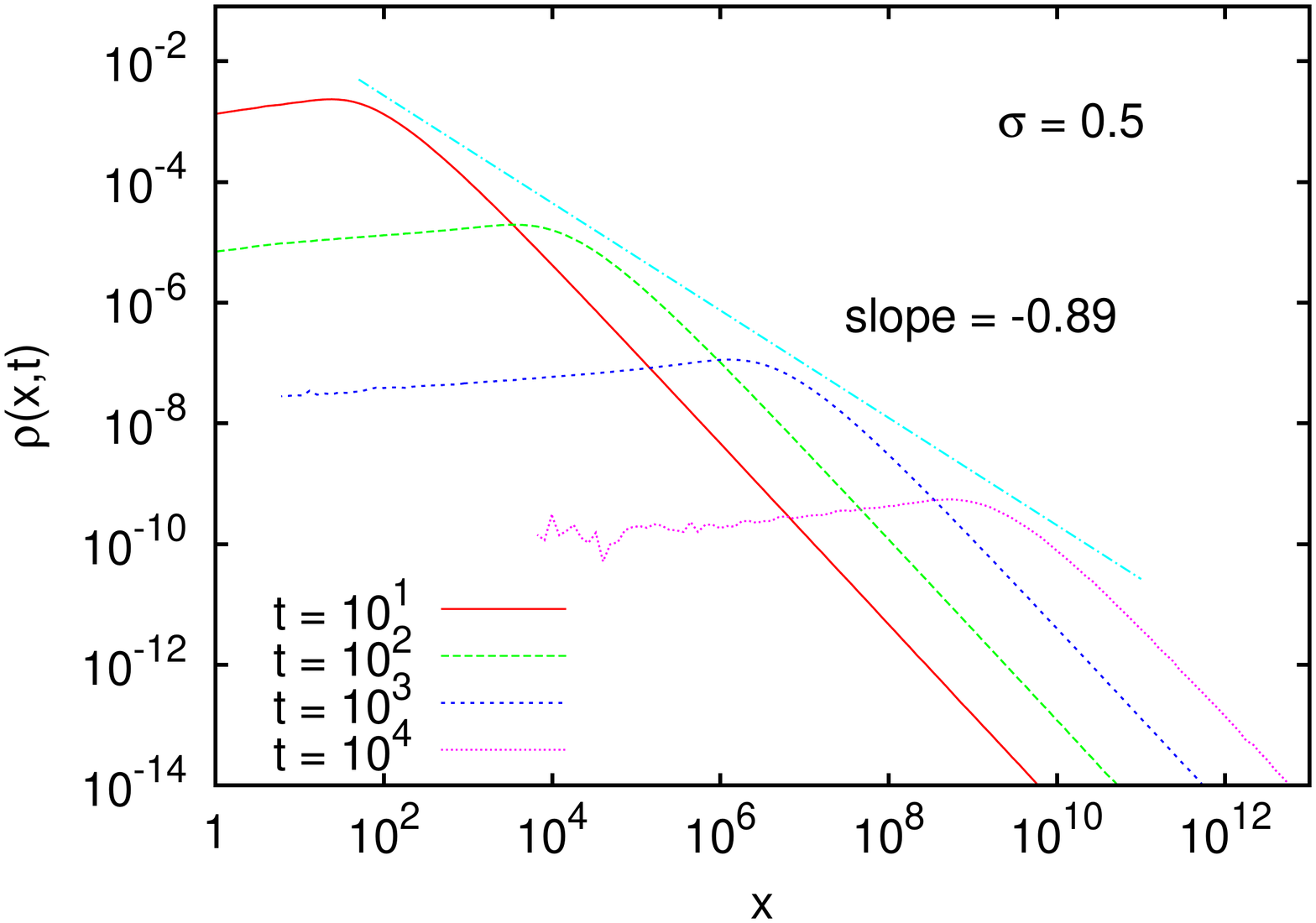}
\caption{(Color online) Densities of active sites for critical epidemics with $\sigma=1/2$, 
for $t = 10, 10^2, \ldots 10^4$. The straight line has the slope predicted by Eq.~(\ref{crit-rhomax}).}
  \label{fig-act-crit}
\end{figure}

As in the case $\sigma=1$ discussed in the previous section, we can also measure 
$z$ directly by measuring the average logarithmic distance of active sites from the 
origin, 
\be
   X_{\rm act}(t) \equiv \exp[\langle \rho(x,t)\; \ln|x| \rangle] \sim t^z\;. \label{defX}
\ee
Values obtained in this way are shown in Fig.~\ref{fig-z-crit}. In this figure we also show the 
values predicted by Eq.~(\ref{z-eta}), finding perfect agreement.

Using Eqs.~(\ref{RPhi}) and (\ref{xit}) we obtain for the density of active sites
\be
   \rho(x,t) = \frac{x^{\sigma-1}}{t}\; \Psi(x/\xi(t))\;,
\ee
from which we find that $\rho(x,t)$ for fixed $t$ has a peak at $x\approx \xi(t)$ with a 
height
\be
   \rho_{\rm max}(t) \sim t^{-1-(1-\sigma)z} \sim \xi(t)^{\sigma-1-1/z}\;. \label{crit-rhomax}
\ee
This prediction is numerically verified for $\sigma=0.5$ in Fig.~\ref{fig-act-crit}.

Finally, we can relate the exponents $\eta$ and $\delta$ by a hyperscaling relation
as follows: Let us consider the growth of two clusters, one starting at position 0 and
the other at $x$, for some large time $t$. If $x=\xi(t)$, the probability that they 
overlap (i.e. have at least one site in common) is of order 1, provided both 
of them have survived up to $t$. Thus, up to constants of order one, 
\be
   R(x,t) \sim \xi(t)^{1-\sigma} \approx [P(t)]^2    \qquad {\rm for} \;\;\; x=\xi(t)
\ee
or 
\be
    \xi(t)^{\sigma-1} \sim t^{(\sigma-1)z} \sim t^{-2\delta}\;,
\ee
which gives \cite{Janssen99,Linder}
\be
    \delta=\frac{1}{2}(1-\sigma)z = \frac{1}{2\sigma}{(1-\sigma)(1+\eta)}\;.  \label{delta-eta}
\ee

Like other hyperscaling relations, it only holds up to the critical dimension,
which in the present case means that it holds for $\sigma\geq 1/3$. For $\sigma<1/3$
the clusters are so sparse that they overlap with probability $<O(1)$. Again, this 
prediction is satisfied for all tested values of $\sigma$. 

\begin{figure}
\includegraphics[width=0.55\textwidth]{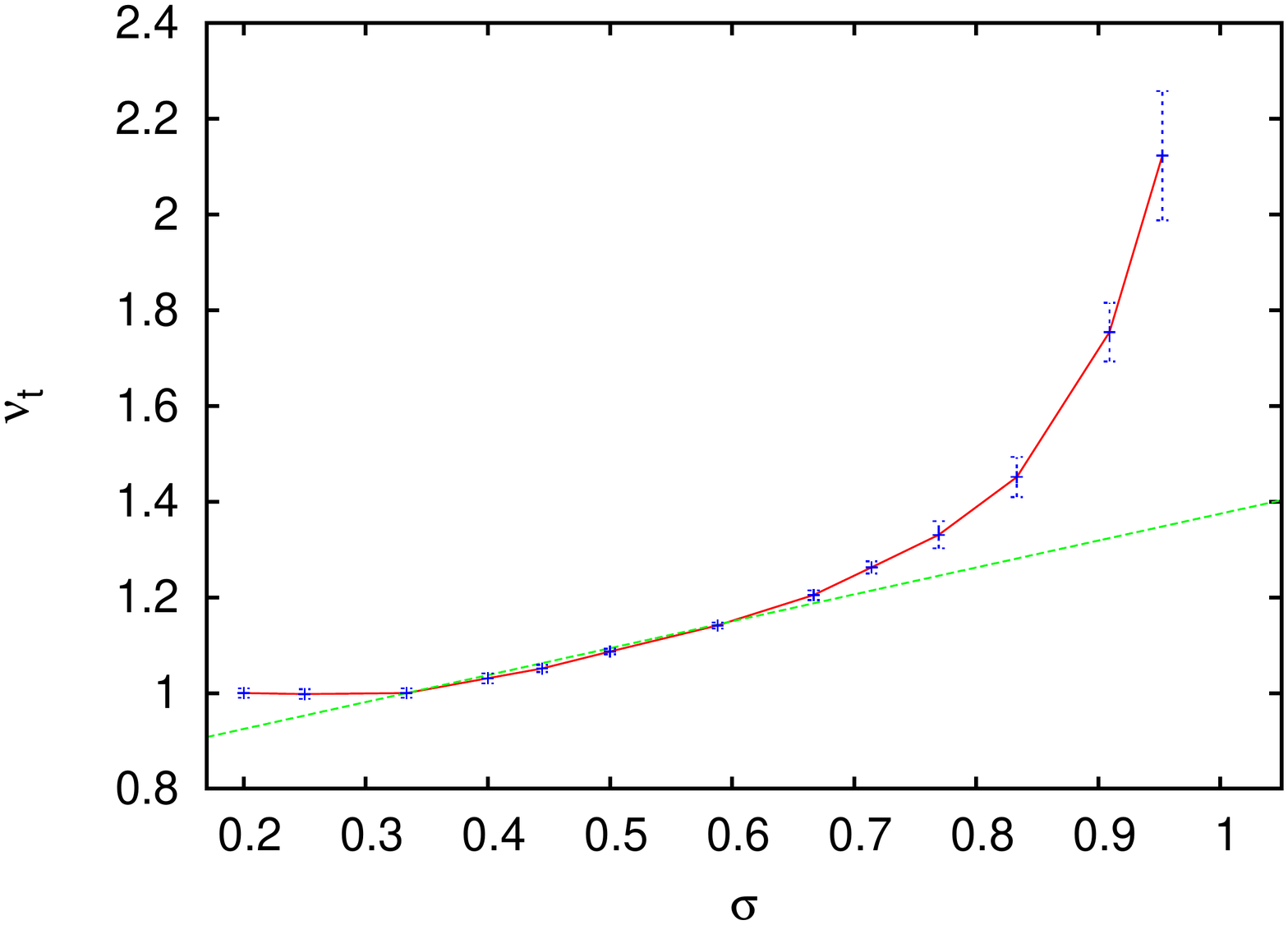}
\caption{(Color online) Correlation time exponent $\nu_t$ against $\sigma$, obtained
by data collapse for the number of infected sites. The straight line is the prediction
if the the $\epsilon$-expansion \cite{Janssen99}. For $\sigma < 1/3$, one has the
mean field prediction $\nu_t=1$. }
  \label{fig-nu-crit}
\end{figure}

As in other critical phenomena, there is one other independent critical exponent
that governs the behavior for $k_{\rm out} \neq k_c$ but $k_{\rm out} \to k_c$.
Traditionally, for percolation with local contacts this can be either the order
parameter exponent $\beta$ or the correlation length exponent $\nu$ (for definitions
and relations between them, see e.g. \cite{Stauffer} or \cite{Linder}. We use
here the correlation time exponent $\nu_t$, which is informally defined via
the scaling of a characteristic time scale $\tau \sim |k_{\rm out} - k_c|^{\nu_t}$.
Using this together with the FTS ansatz Eq.~(\ref{nut-scaling}), we can then 
determine numerical values of $\nu_t$ by plotting $t^\delta n(t)$ versus
$(k_{\rm out} - k_c)t^y$ for different trial values of $y$. 
Data collapse should occur for large $t$ and finite $(k_{\rm out} - k_c)t^y$, when
$y=1/\nu_t$. Values obtained in this way are shown in Fig.~\ref{fig-nu-crit}. We
again see that the mean field prediction $\nu_t=1$ for $\sigma<1/3$ is satisfied,
and that the $\epsilon$-expansion \cite{Janssen99} is correct for $\sigma$ slightly
larger than 1/3. For $\sigma\to 1$ one finds that $\nu_t$ diverges, which is
in agreement with the prediction \cite{Thouless,Cardy,Aizenman} than the transition is
of BKT type for $\sigma=1$.

As a last remark we should point out that the theoretical discussion of the present 
section applies only to infinite systems. For critical phenomena with short range 
interactions, the correlation length $\xi$ describes both what happens in finite 
systems and, in infinite systems, at finite times or finite distances from the 
critical point. In the present case, due to the definition of $\xi$ via a geometric 
average, it is less clear how it relates to the finite system size behavior.

\section{Supercritical epidemics with $0 < \sigma < 1$}   \label{supercrit}

In contrast to the critical case and to the case with $\sigma=1$, where the epidemics
are described by scaling laws, the situation seems much less clear for 
supercritical epidemics with $0 < \sigma < 1$. 

In this case mean field theory predicts for both models exponential growth of 
$N(t)$ and of the spatial extent 
\cite{Mancinelli,Castillo}. The reason is very simple. For $\sigma < 1$ the 
wave of infection propagates, in mean field theory, like a `pulled' \cite{Ebert-2000}
front, i.e. the growth of the cluster is mainly controlled by its most 
advanced `avant garde' (for $\sigma>1$, in contrast, we have seen that the front 
is `pushed' by the region where the density is large). In this region ahead of the main
front the 
density is very small, and thus saturation effects are negligible. Practically 
every attempted infection also succeeds, and thus the density increases exponentially 
with time as 
\be
    R\sim k_{\rm out}^t \qquad (\mbox{ for fixed large}\;\;\; x) 
\ee
Together with the spatial power law $R \sim x^{-1-\sigma}$ this means 
that an effective front position $x_{\rm front}(t)$, defined by 
\be
   R(x_{\rm front}(t),t) = const\;,
\ee
must advance exponentially, like \cite{Mancinelli}
\be
   x_{\rm front}(t) \sim k_{\rm out}^{t/(1+\sigma)} \;.  \label{X_mf}
\ee
As we have seen in Sec.~II, this is {\it not} supported by our data.

The argument leading to Eq.~(\ref{X_mf}) was criticized in \cite{Brockmann}, who 
argued that the advance of the front should be linear in time. Unfortunately, 
this is not supported by the data either, and it is easy to see why. 
In \cite{Brockmann} it was assumed that new infections are not successful, if 
they appear in regions with extremely small density. This would be correct, if 
we had assumed {\it cooperativity} (or `synergy') in the infection \cite{Bizhani,Perez-Reche}.
But this was not assumed in \cite{Mancinelli,Castillo}, nor was it stated
explicitly in \cite{Brockmann}. It is also not assumed in the present paper.

We claim that the problem is, instead, a break-down of mean field theory. It 
is true that the front is pulled, and it is true that the {\it average} 
density, averaged over the entire ensemble, is small in the tail of the front. 
But as soon as a site far ahead of the previous front (i.e., in a region 
with very small density) is infected, it will generate its own little `colony'
and create locally a spot with high density. Thus, in spite of the very small
average densities, there will always be non-negligible saturation effects 
due to the not-so-small {\it actual} densities.

\begin{figure}
\includegraphics[width=0.55\textwidth]{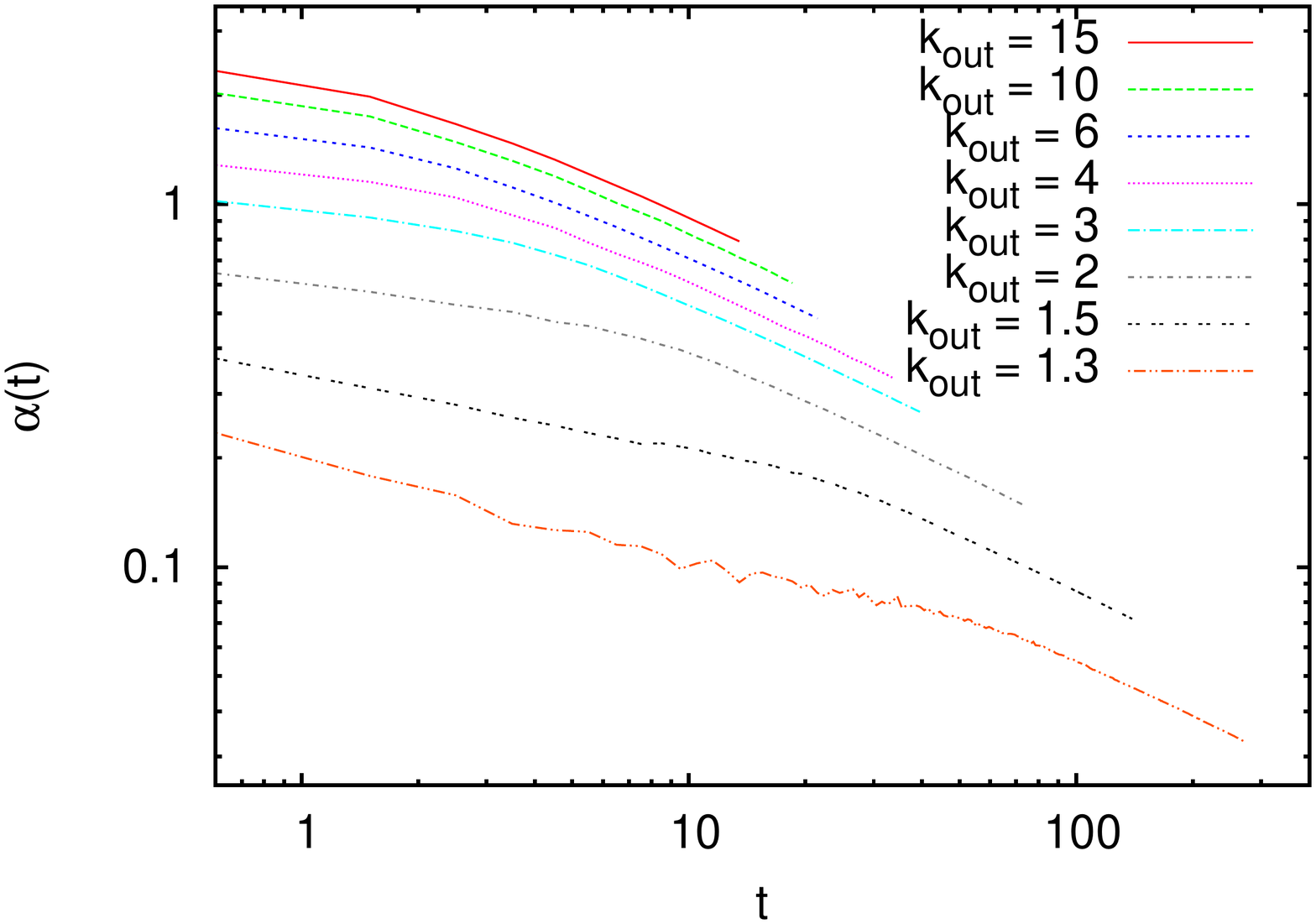}
\caption{(Color online) Log-log plot of growth rates (as defined in Eq.~(\ref{gr-rate}))for
 $\sigma=0.5$, and for different values of $k_{\rm out}$. In all cases, the measured
growth rates are much smaller than those predicted by Eq.~(\ref{gr-mf}), except for 
very small $t$.}
  \label{fig-growthrates}
\end{figure}

\begin{figure}
\includegraphics[width=0.55\textwidth]{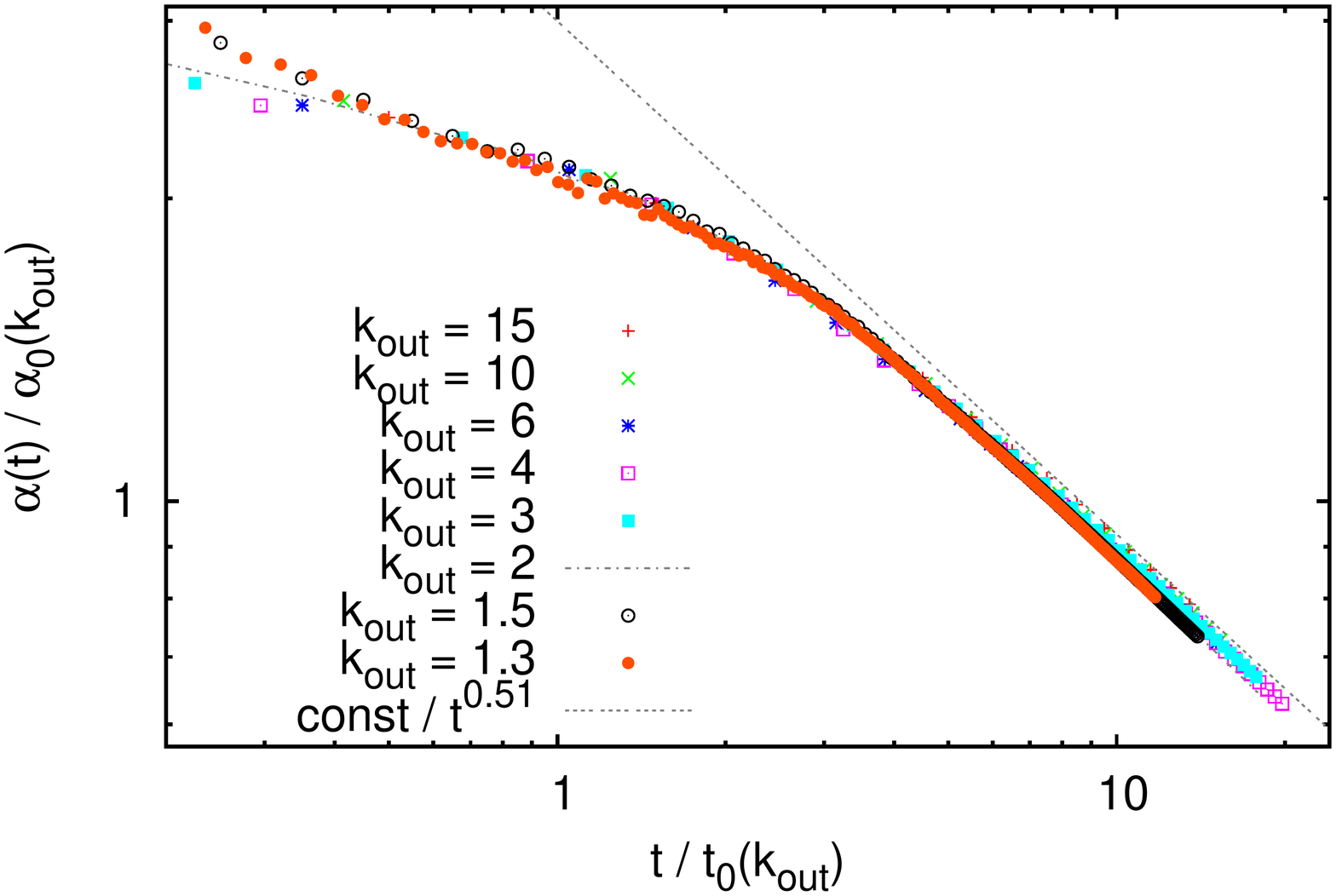}
\caption{(Color online) Log-log plot of re-scaled growth rates against re-scaled time.
The straight line corresponds to $\gamma=0.49$.}
  \label{fig-growthrates-collapse}
\end{figure}

Let us define a time dependent growth rate $\alpha(t)$ by
\be
   \alpha(t) = \ln \left[\frac{n(t+1/2)}{n(t-1/2)}\right].     \label{gr-rate}
\ee
Growth rates for $\sigma=0.5$ and different values of $k_{\rm out}$ are shown 
in Fig.~\ref{fig-growthrates}. According to Eq.~(\ref{X_mf}) we should 
expect
\be
   \alpha(t) = \alpha_{\rm mf}(t) \equiv \frac{\ln k_{\rm out}}{1+\sigma}\;,    \label{gr-mf}
\ee
but we see that the measured $\alpha$ drops, for all tested values of $k_{\rm out}$,
far below this mean field value. Instead, it seems that $\alpha(t)$ decreases
for large $t$ like a power 
\be
   \alpha(t) \sim \alpha_0 t^{\gamma-1},    \label{ag}
\ee
where $\alpha_0$ depends on $k_{\rm out}$, but $\gamma$ is independent of 
$k_{\rm out}$. In particular for $\sigma=0.5$, we obtain $\gamma=0.49(2)$.

This means that for large $t$ 
\be
    n(t) \sim e^{bt^{\gamma}}
\ee
with $b=\alpha_0/\gamma$, e.g. for $\sigma=0.5$ we have a stretched exponential
with exponent $0.49(2)$. This is particularly evident if we collapse these data 
onto a single curve by rescaling $t$ and $\alpha$ by arbitrary functions of 
$k_{\rm out}$, see Fig.~\ref{fig-growthrates-collapse}.

The value $\gamma=0.49$ is to be compared with Sec. II,
where we obtained an exponent $0.59(1)$ by a straightforward fit. The 
discrepancy between these two estimates results from systematic deviations
from a pure stretched exponential, which are indeed visible in both 
Figs.~\ref{n_t-s15} and \ref{fig-growthrates}. In both cases the curves bend 
downward (instead of being straight as for clean stretched exponentials), 
indicating that a naive fit overestimates the growth. Thus it seems likely that 
the present estimate obtained via the growth rate is more reliable. In addition, 
we made similar fits with $n(t)$ replaced by $N(t)$. Both should satisfy asymptotically 
the same stretched exponential, but with different power prefactors. The 
estimate of $\gamma$ via the growth rate seems more robust than the direct estimate
of Sec.~II. Finally, we estimated by both methods (direct fit \& growth rate) the 
exponents for trial functions of the type $f(x) = x^a \exp(bt^\gamma)$ with various
(positive and negative) prefactor powers. In most cases the growth rate method gave 
better results.

\begin{figure}
\includegraphics[width=0.55\textwidth]{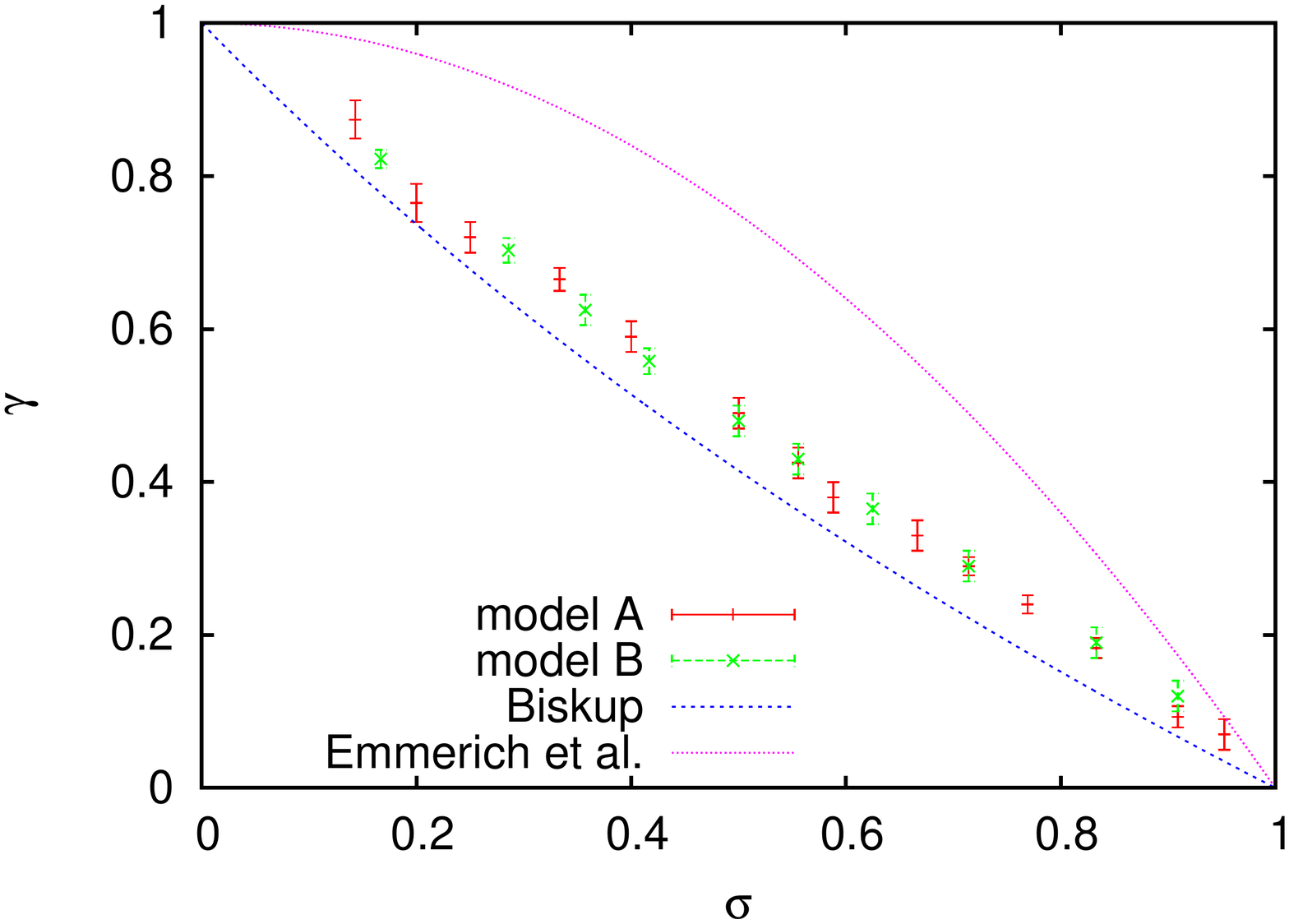}
\caption{(Color online) Stretched exponential powers as a function of $\sigma$ for 
   models (A) and (B). The continuous curves are the predictions of Biskup \cite{Biskup-2004}
   (lower curve) and Emmerich {\it et al.} \cite{Emmerich} (upper curve).}
  \label{fig-gamma}
\end{figure}

For other values of $\sigma$, very similar results were found. In particular,
 the exponent $\gamma$ is also independent of 
$k_{\rm out}$ within the estimated errors \footnote{This is very clearly seen
for $\sigma\approx 1$, where also Eq.~(\ref{ag}) holds already for small $t$. 
For $\sigma\approx 0$ Eq.~(\ref{ag}) holds only for very large $t$, and the 
fitted values of $\gamma$ decrease weakly with $k_{\rm out}$. We interpret
this as a finite-size artifact.}, suggesting that it really 
depends only on $\sigma$. The values obtained in this way for both models are 
shown in Fig.~\ref{fig-gamma}. We see that both models give very similar 
results. Both are consistent, within the errors, with $\gamma = 1-\sigma$, but 
it seems that this is not the correct behavior. A percolation model very similar to 
the static part of model (A) was studied rigorously by Biskup \cite{Biskup-2004,Biskup-2009}. 
He obtained results for the graph diameter of clusters embedded in large but finite
lattices \cite{Biskup-2009} and for the average graph distance between nodes 
with large Euclidean distance on infinite lattices \cite{Biskup-2004}. Neither of 
this is precisely what is measured in the present work, but his results strongly suggest 
that
\be
    \gamma = \frac{\ln 2}{\ln\frac{2}{1+\sigma}}.
\ee
This prediction is also shown in Fig.~\ref{fig-gamma}. It agrees reasonably well with
our simulations. The simulations are systematically too high, indicating that we have 
still substantial finite cluster size corrections. Notice that naive fits like 
Fig.~\ref{n_t-s15} would give even larger estimates.

The graph diameters of graphs embedded in finite lattices were also measured 
by \cite{Kosmidis,Emmerich}. Unfortunately, these authors used a rather complicated 
algorithm which prevented them from using very large lattices and from obtaining 
high statistics. It also introduced particularly large finite lattice corrections, 
and the data were analyzed by fitting stretched exponentials directly via 
plots like Fig.~\ref{n_t-s15}. It  is presumably for these reasons 
that \cite{Kosmidis,Emmerich} obtain $\gamma=1-\sigma^2$ (in our notation), 
which is clearly incompatible with our data and with the prediction 
of \cite{Biskup-2004,Biskup-2009} (see Fig.~\ref{fig-gamma})
\footnote{I might also add that, based on the same model, it is claimed
in \cite{Daqing,Emmerich} that such networks embedded in $d$ Euclidean dimensions
have in general fractal dimension (as measured via the Euclidean distance) $d_f > d$.
This is of course impossible, as it would violate one of the most basic properties
of any fractal \cite{Falconer}.}.

Figure~\ref{immune-s15} suggests a scaling ansatz 
\be
   R(x,t) = \phi(x/\xi(t))     \label{Rphi_super}
\ee
similar to Eq.~(\ref{Rphi}), but with $\xi(t)$ being a stretched exponential,
\be
   \xi(t) \sim N(t) \sim \exp(bt^\gamma)\;,    \label{xi_t-super}
\ee
instead of a power law. From this ansatz follows
\be
   \rho(x,t) = t^{\gamma-1} \psi(x/\xi(t))
\ee
where $\psi(z) \propto -z^{-1} \phi'(z)$.
This means in particular that $\rho_{\rm max}(t)$, defined in Eq.~(\ref{eta-eta}),
decreases like a power of $t$,
\be
   \rho_{\rm max}(t) = t^{\gamma-1}  \;.          \label{rho-gamma}
\ee
This is easy to understand. As we pointed out already at the beginning of this 
section, the effective growth rate decays
as $ t^{\gamma-1} $, because new attempted infections are increasingly more 
likely to target sites that are no longer susceptible. But there it applied 
to the `front of the front', while here
we see that it applies also to the `core of the front' where most of the mass 
growth occurs. We can interpret Eq.~(\ref{rho-gamma}) also as a manifestation 
of a weak sort of `self-organized criticality' \cite{Bak} in the sense that 
the speed of growth is such that the density at the front $x\approx \xi(t)$ 
converges exactly to its critical value. 

Equations~(\ref{rho-gamma}) and (\ref{xi_t-super}) together imply the remarkable
relation 
\be
   \frac{\log \xi(t)}{\rho_{\rm max}(t)} \sim t\;,  \label{xi-rho}
\ee
which should hold for any value of $\sigma \in [0,1)$ and does not involve 
$\sigma$ or $\gamma$ explicitly.

\begin{figure}
\includegraphics[width=0.55\textwidth]{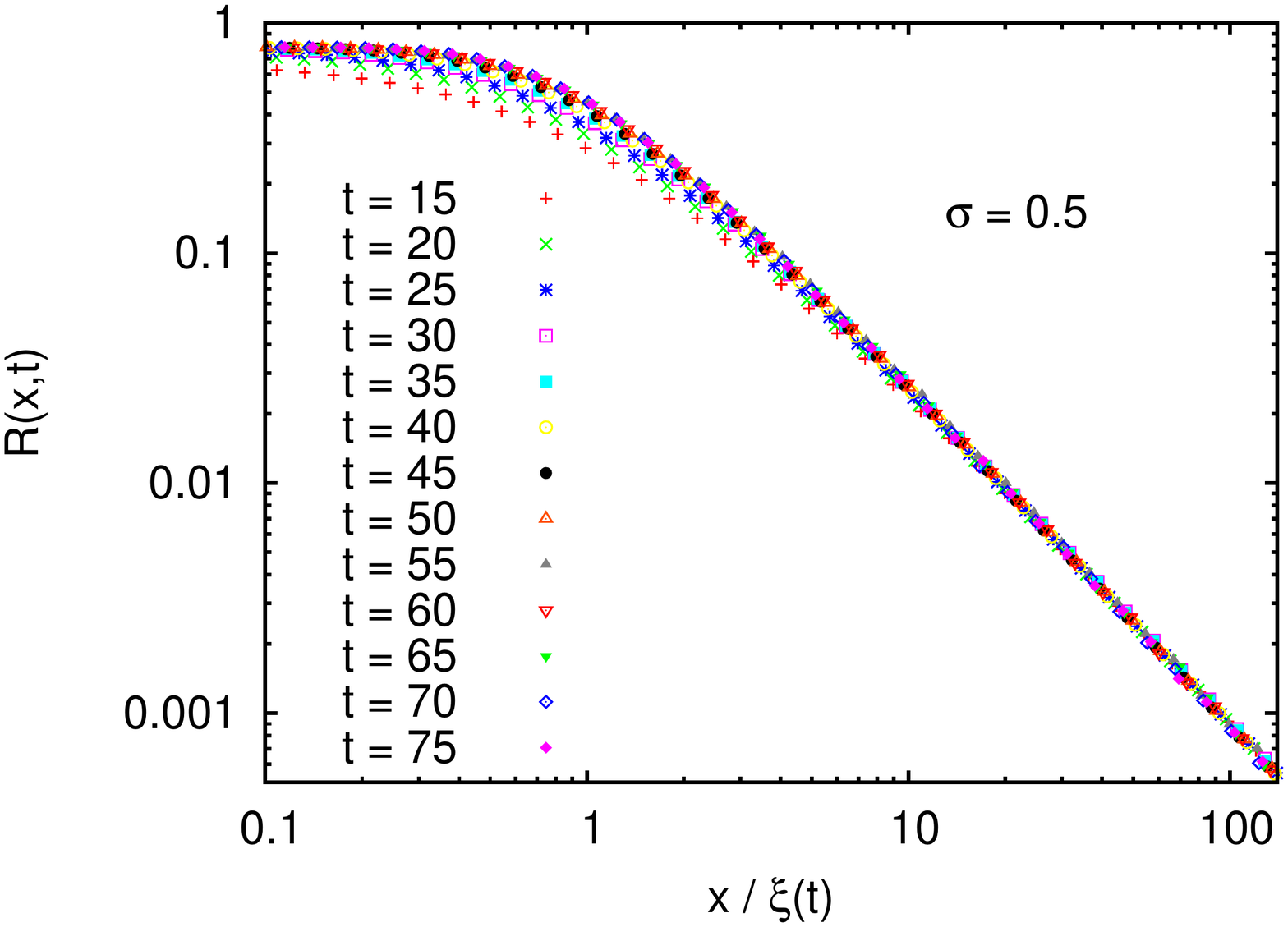}
\caption{(Color online) Collapse plot for the data (for $t\geq 15$ only) shown 
in Fig.~\ref{immune-s15}. We see a perfect data collapse for $t\geq 50$.}
  \label{fig-Rphi_super-collapse}
\end{figure}

\begin{figure}
\includegraphics[width=0.55\textwidth]{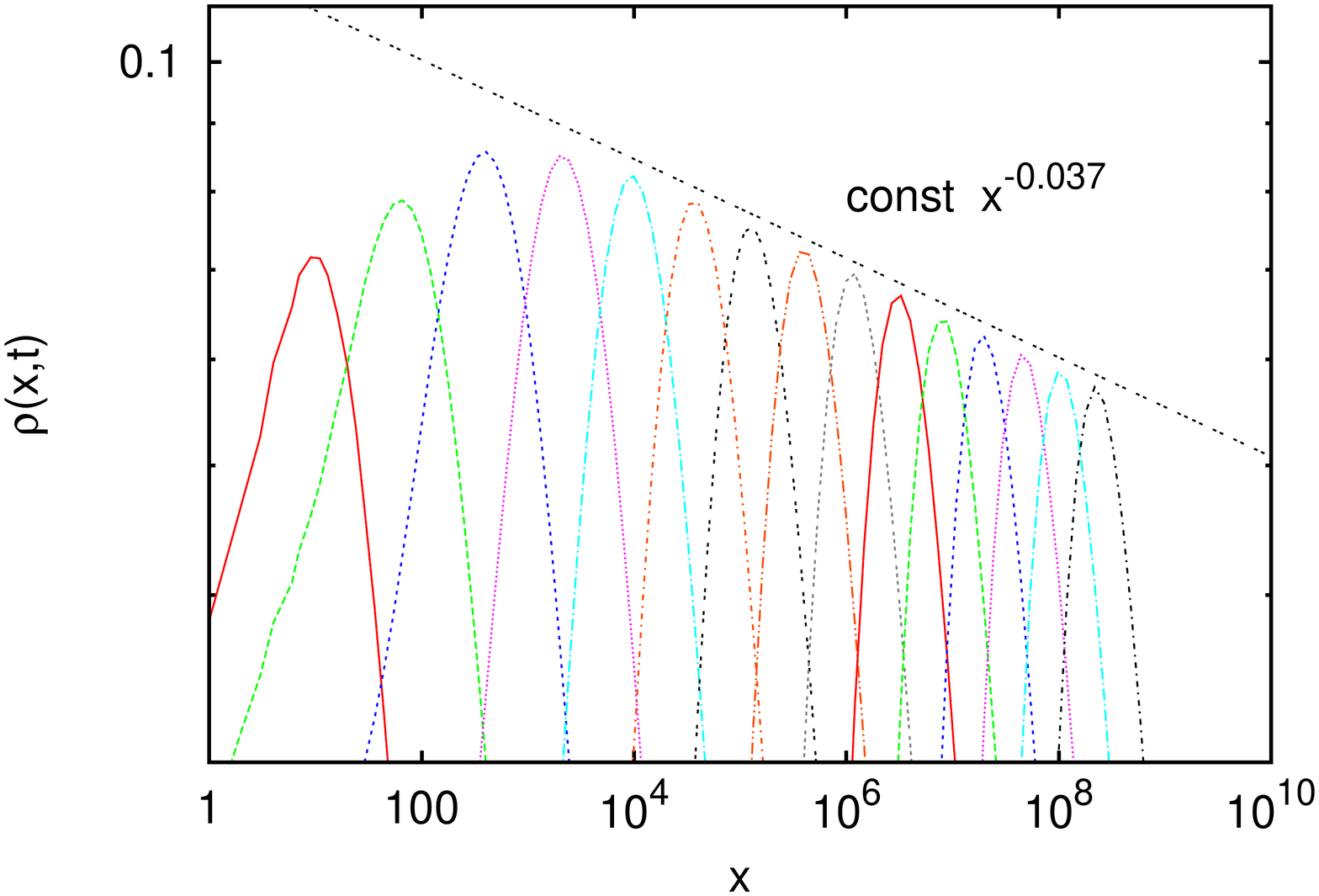}
\caption{(Color online) Enlargement of part of Fig.~\ref{active-s15}. It shows
that the peak heights decrease with $t$, if $t>20$. For very large $t$, where 
Fig.~\ref{fig-Rphi_super-collapse} shows a data collapse, this decrease is 
compatible with a power law $\rho_{\rm max} \sim \xi^{-\beta}$ with very small
$\beta$, but we shall argue that this dependence is actually logarithmic.}
  \label{fig-rho-gamma}
\end{figure}

\begin{figure}
\includegraphics[width=0.55\textwidth]{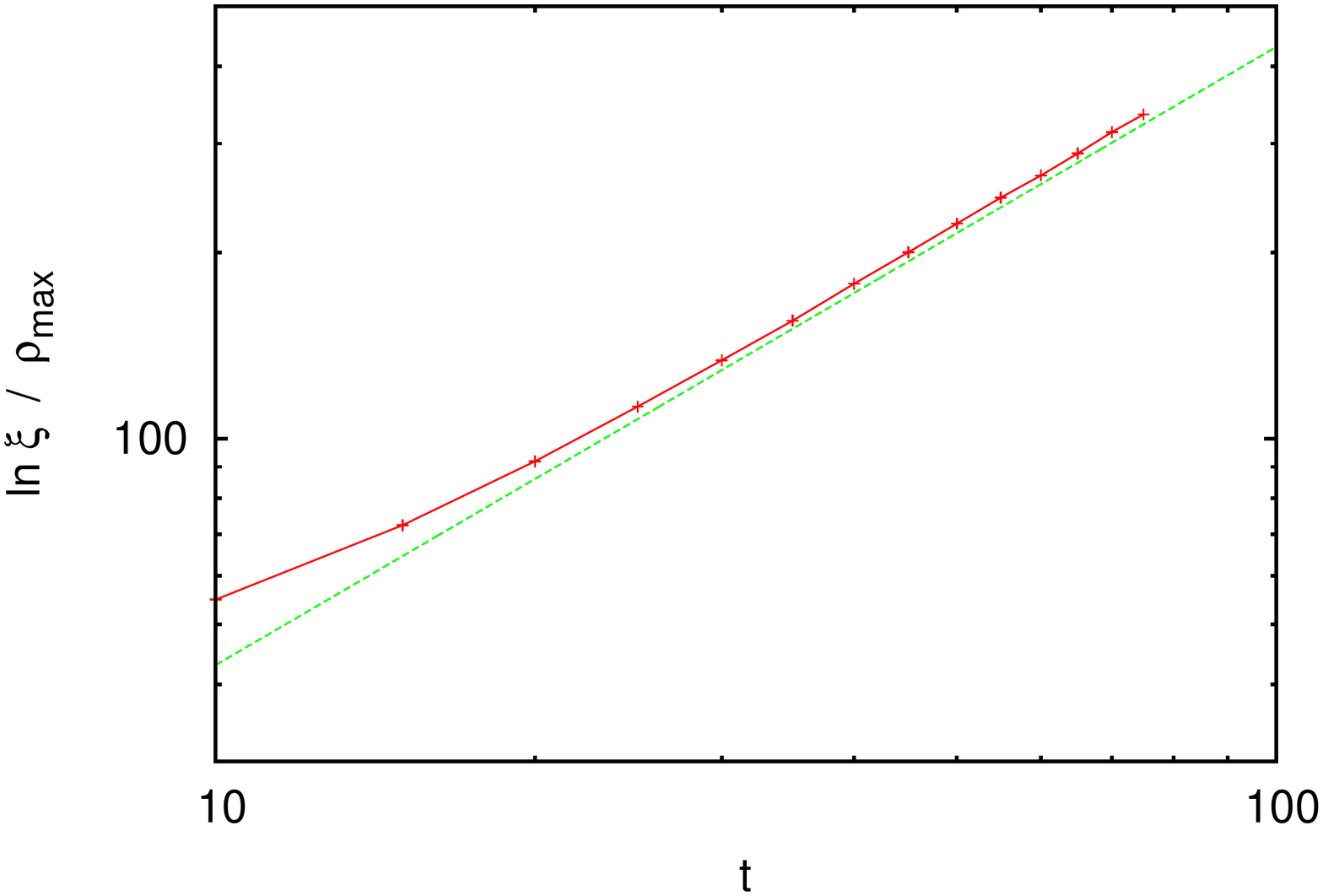}
\caption{(Color online) Log-log plot of $\rho_{\rm max}, \ln \xi$, and of
$\ln \xi/\rho_{\rm max}$ against $t$. According to Eq.~(\ref{xi-rho}) the latter 
should be proportional to $t$ in the large $t$ limit, which according to 
Fig.~\ref{fig-Rphi_super-collapse} should be reached for $t\approx 50$. The 
straight line has slope 1.}
  \label{fig-xi-rho}
\end{figure}

Equations (\ref{Rphi_super}) to (\ref{xi-rho}) are tested numerically,
again for $\sigma=0.5$ and $k_{\rm out}=2$, in Figs. \ref{fig-Rphi_super-collapse}
to \ref{fig-xi-rho}. In Fig.~\ref{fig-Rphi_super-collapse} we see a 
perfect data collapse for sufficiently large $t$, while we see in 
Fig.~\ref{fig-rho-gamma} that 
\be
   \rho_{\rm max} = \xi^{-\beta}
\ee
with $\beta = 0.037$ seems to fit the large time asymptotics.
Together with Eq.~(\ref{xi_t-super}), this would however imply that $\rho_{\rm max}(t)$
decreases with $t$ faster than a power, which is incompatible with Eq.~(\ref{rho-gamma}).
Therefore we show in Fig.~\ref{fig-xi-rho} the ratio $\frac{\log \xi(t)}{\rho_{\rm max}(t)}$
as a function of $t$, in order to test Eq.~(\ref{xi-rho}). Here, $\xi(t)$ is as obtained
in the collapse shown in Fig.~\ref{fig-Rphi_super-collapse}. This defines $\xi(t)$ only
up to a constant. This constant (used, by the way, also in Fig.~\ref{fig-Rphi_super-collapse})
is fixed so that a straight line is obtained in Fig.~\ref{fig-xi-rho}. We see that
the data follow indeed a nice linear relationship, showing that at least the entire
scheme is internally consistent.

\begin{figure}
\includegraphics[width=0.55\textwidth]{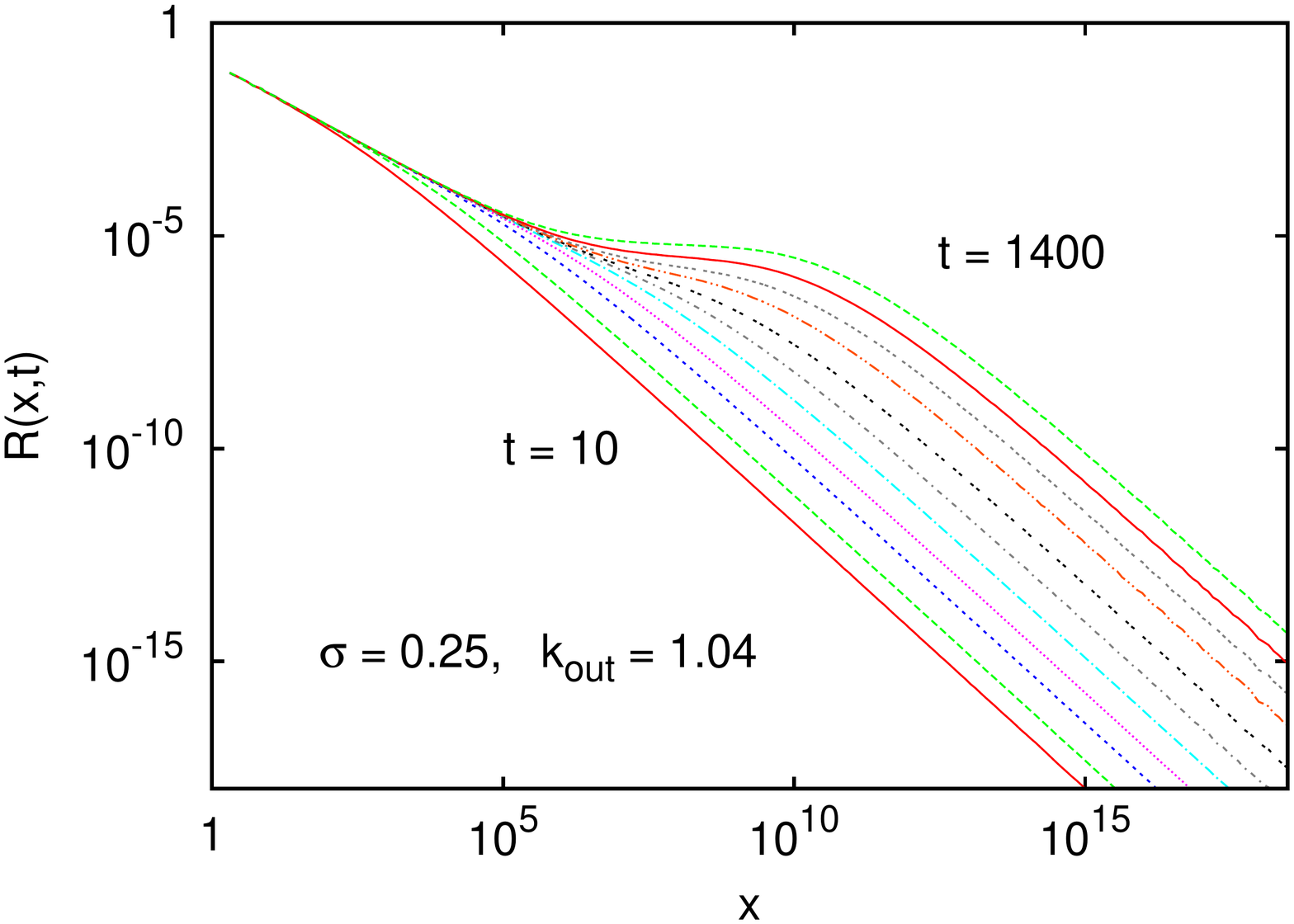}
\caption{(Color online) Log-log plot of $R(x,t)$ versus $x$, for $t=10,20,50,100,
200,350,550,800,1000,1200$,and 1400. Control parameters are $\sigma=0.25$ and 
$k_{\rm out}=0.04$. The latter is ca 20\% above $k_c$. }
  \label{fig-immune-4_t}
\end{figure}

Finally, before leaving this section, let us point out that the scaling relation 
(\ref{Rphi_super}) sets in very late when $k_{\rm out} \approx k_c$, in particular 
when $\sigma <1/3$. In this range of $\sigma$, mean field theory hold for 
$k_{\rm out} = k_c$, but it does not hold for any $k_{\rm out} > k_c$. Thus, the 
cross-over from mean-field to `semi-local' behavior has to happen in a very narrow 
region with $k_{\rm out} - k_c \ll 1$. We illustrate this with Fig.~\ref{fig-immune-4_t},
where we show $R(x,t)$ for $\sigma=0.25$ and $k_{\rm out} = 0.04$ versus $x$ for 
different values of $t$, in a way completely analogous to Fig.~3 (where $\sigma = 0.5, 
k_{\rm out}=2$). Instead of the
structureless curves in Fig.~3, we see now that for short times the system behaves
as if it were critical and mean field. For $t> 200$ it becomes clear that the 
process is supercritical, but newly infected sites are very far from the seed. It is 
only for $t>1000$ that the density starts to grow again appreciably at intermediate
distances $x \approx 10^6$, in order to reach finally its asymptotic value $O(1)$.

\section{Discussion and Conclusions}

Networks embedded in space with connections which preferentially link close 
neighbors but have also non-vanishing chances to link nodes far apart have numerous
applications, from biology to social sciences. Very early it was already proposed
to model the link length distribution by power laws \cite{Mollison,Grassberger86}.
In the present paper we generate such networks by epidemic processes with power
behaved distributions for contacts, i.e. infections occur over distances $x$ whose 
probability decays as $P(x) \sim 1/x^{1+\sigma}$.

While the case of two spatial dimensions will be treated in a forthcoming paper
\cite{Grassberger2013}, we restricted ourselves here to $d=1$. This is of course 
less interesting from the point of view of applications, but it allows much more 
detailed and precise analysis. Our results concern mainly three different regimes:\\
(i) critical epidemics with $0< \sigma < 1$,\\
(ii) epidemics with $\sigma =1$,\\
(iii) supercritical epidemics with $0< \sigma <1$.\\
In all three cases we obtain significant new results, by using simulations on
unprecedentedly large sizes. Using hashing, the lattices we used have 
$2^{64}\approx 1.8\times 10^{19}$ sites. Such large lattices are needed in 
order to avoid finite size effects, in view of infections that can spread 
over billions of sites in one time step.

For critical epidemics we verify predictions from field theory and show that 
there is one independent critical exponent less than for critical epidemics
with short range contacts in $\geq 2$ dimensions (in $d=1$, epidemics with 
short range contacts die out). For supercritical epidemics with $\sigma =1$
we verify old predictions based on the Fortuin-Kasteleyn connection between 
percolation and Potts models. And for supercritical epidemics with $0< \sigma <1$
we verify predictions \cite{Biskup-2004,Biskup-2009} that they lead to `medium-size 
world' networks, i.e. to networks that are not `small world' in the sense that 
their size grows exponentially with their graph diameter, but which are also
not fractal in the sense that this mass grows like a power. Instead, it 
grows like a stretched exponential. Related to this is the observation that 
supercritical epidemics with $0<\sigma <1$ spread neither with fixed nor 
with exponentially increasing velocity, in contradiction to previous claims.

In all three cases we find several new scaling laws that are strongly 
suggested numerically and which we show in some cases to satisfy non-trivial
consistency relations, but for which we do not give theoretical derivations.
The need to provide these proofs is one of the main open problems. 


In particular, we verify numerically that percolation with $P(x) \sim 1/x^2$ is
discontinuous in one dimension, as proven in \cite{Aizenman,Aizenman88}. It 
is intriguing that percolation is also discontinuous in the model of Boettcher 
{\it et al.} \cite{Boettcher-2012}, which can also be understood as a $1-d$ 
lattice with additional long range links whose number decreases with $x$ in 
the same way. The main difference between the two models is that the long 
range links in the Boettcher model are less random (their lengths can only be 
a power of 2, and they
attach only to selected sites). It would be of interest to check whether the 
transition in the Boettcher model is also BKT-like as regards the increase of 
mass of supercritical clusters with their graph diameter.
 
Finally, a last problem which we left open is finite size behavior. Our strategy 
was to use lattice sizes which are big enough so that finite (lattice-)size
effects can be safely neglected. In view of the interplay between the 
length scales set by the lattice size, the critical correlation length,
and the large contact distances, we may expect finite size effects to be 
not as simple as in conventional finite size scaling for critical phenomena 
with short range interactions.

\section*{Acknowledgements}

For very helpful discussions I want to thank Aicko Schumann and Deepak Dhar.
I also want to thank the latter for the kind hospitality at the Tata Institute
of Fundamental Research in Mumbai, where part of this work was done, and for 
carefully reading the manuscript. The work was begun at the Complexity Science 
Group at the Universality of Calgary, which I also want to thank for generous 
grant of computer time. Finally my thanks go to Haye Hinrichsen and Hans-Karl 
Janssen for illuminating correspondence, and to an anonymous referee for pointing 
out the similarity with Ref.~\cite{Boettcher-2012}.

\bibliography{mm}

\end{document}